\documentclass[11pt,a4paper]{article}

\usepackage{amssymb}
\usepackage{graphicx}
\usepackage{bm}

\usepackage{mathrsfs}
\usepackage{cite}

\begin{document}

\title{Three-loop $\beta$-functions and two-loop anomalous dimensions for MSSM regularized by higher covariant derivatives in an arbitrary supersymmetric subtraction scheme}

\author{O.V.Haneychuk${}^a$, V.Yu.Shirokova${}^a$, K.V.Stepanyantz${}^{abc}$ $\vphantom{\Big(}$
\medskip\\
{\small{\em Moscow State University, Faculty of Physics,}}\\
${}^a${\small{\em Department of Theoretical Physics,}}\\
${}^b${\small{\em Department of Quantum Theory and High Energy Physics,}}\\
\medskip
{\small{\em 119991, Moscow, Russia;}}\\
$^c$ {\small{\em Bogoliubov Laboratory of Theoretical Physics, JINR,}}\\
{\small{\em 141980 Dubna, Moscow region, Russia.}}
}

\maketitle

\begin{abstract}
Three-loop $\beta$-functions of the Minimal Supersymmetric Standard Model regularized by higher covariant derivatives are obtained for an arbitrary supersymmetric subtraction scheme. For this purpose we first calculate two-loop anomalous dimensions for all MSSM chiral matter superfields defined in terms of the bare couplings. Then we use the NSVZ equations for the renormalization group functions defined in terms of the bare couplings, which are valid in all orders in the case of using the higher covariant derivative regularization. This gives the three-loop $\beta$-functions defined in terms of the bare couplings. After that, we construct the three-loop $\beta$-functions and the two-loop anomalous dimensions standardly defined in terms of the renormalized couplings for an arbitrary subtraction scheme. As a nontrivial correctness test, we verify that for a certain renormalization prescription the general results reproduce the ones obtained earlier in the $\overline{\mbox{DR}}$ scheme. Also this can be considered as an indepedent confirmation of the $\overline{\mbox{DR}}$ results.
\end{abstract}

\section{Introduction}
\hspace*{\parindent}

Investigation of quantum corrections in various field theory models is very important for understanding nature. For instance, the coincidence of the first few digits in the theoretical and experimental values of the electron anomalous magnetic moment demonstrates that the surrounding world is described by quantum field theory \cite{Peskin:1995ev}, although the existing discrepancies (which remain even after taking into account the results of the 5-loop perturbative calculation in quantum electrodynamics) are the subjects of various theoretical and experimental studies \cite{Aoyama:2019ryr}. The precise measurements of the muon magnetic moment and the comparison of the results with the theoretical predictions can shed light on physics beyond the Standard Model, see, e.g., \cite{Li:2021bbf,Keshavarzi:2021eqa}. The unification of gauge coupling constants (if any) may be considered as an argument in favour of the realization of supersymmetry based models of Grand Unification (see, e.g., \cite{Ellis:1990wk,Amaldi:1991cn,Langacker:1991an}). That is why quantum corrections in ${\cal N}=1$ supersymmetric theories and theories with softly broken supersymmetry are especially interesting \cite{Mihaila:2013wma}. In particular, it is important to know the $\beta$-functions in such theories, which encode the evolution of the running gauge coupling constants. In the most popular $\overline{\mbox{DR}}$ scheme (when a theory is regularized by dimensional reduction \cite{Siegel:1979wq} and divergences are removed by modified minimal subtraction \cite{Bardeen:1978yd}) the $\beta$-function has been calculated in the three- \cite{Avdeev:1981ew,Jack:1996vg,Jack:1996cn,Jack:1998uj} and four-loop \cite{Harlander:2006xq} approximations. In the Minimal Supersymmetric Standard Model (MSSM)  the three-loop expressions for the $\beta$-functions in the $\overline{\mbox{DR}}$ scheme have been found in \cite{Jack:2004ch}. However, from the theoretical point of view, the $\overline{\mbox{DR}}$ scheme is not the best renormalization prescription. Really, it is known that dimensional reduction is not mathematically consistent \cite{Siegel:1980qs} and can break supersymmetry in higher loops \cite{Avdeev:1981vf,Avdeev:1982np,Avdeev:1982xy}. Moreover, the renormalization group functions (RGFs) in supersymmetric theories in certain subtraction schemes satisfy the NSVZ equation \cite{Novikov:1983uc,Jones:1983ip,Novikov:1985rd,Shifman:1986zi}, which relates the $\beta$-function to the anomalous dimension of the matter superfields. The renormalization prescriptions for which the NSVZ equation is valid are usually called the NSVZ schemes. According to \cite{Jack:1996vg,Jack:1996cn,Jack:1998uj}, the $\overline{\mbox{DR}}$ scheme is not an NSVZ scheme. However, although the NSVZ relation is not satisfied for the $\overline{\mbox{DR}}$ prescription, it can be restored by a special finite renormalization compatible with a structure of quantum corrections. The MOM scheme is also not NSVZ \cite{Kataev:2013csa,Kataev:2014gxa}. (However, for ${\cal N}=1$ supersymmetric electrodynamics in the on-shell scheme the NSVZ relation is valid in all orders \cite{Kataev:2019olb}.)

Some all-loop NSVZ schemes can be constructed with the help of the higher covariant derivative regularization \cite{Slavnov:1971aw,Slavnov:1972sq} in the supersymmetric formulation \cite{Krivoshchekov:1978xg,West:1985jx}. Note that this regularization includes the insertion of the Pauli--Villars determinants into the generating functional for removing residual one-loop divergences \cite{Slavnov:1977zf}, see \cite{Aleshin:2016yvj,Kazantsev:2017fdc} for the generalization to the supersymmetric case. Since this regularization is consistent and formulated in the dimension $D=4$, it is more preferable for supersymmetric theories than dimensional reduction. (The detailed comparisons of various regularizations can be found in \cite{Jack:1997sr,Gnendiger:2017pys}.) With the higher covariant derivative regularization RGFs defined in terms of the bare couplings (which are scheme-independent for a fixed regularization) satisfy the NSVZ relation. For usual RGFs (defined in terms of the renormalized couplings) some NSVZ schemes can be obtained by the HD+MSL prescription, when for a theory regularized by Higher Derivatives divergences are removed by Minimal Subtractions of Logarithms. By other words, the renormalization constants should include only powers of $\ln\Lambda/\mu$, where $\Lambda$ is the dimensionful regularization parameter and $\mu$ is the renormalization point. The proof of these facts for theories with simple gauge groups has been done in Refs. \cite{Stepanyantz:2016gtk,Stepanyantz:2019ihw,Stepanyantz:2020uke}. (For the Abelian case it was constructed earlier in Refs. \cite{Stepanyantz:2011jy,Kataev:2013eta}.) Similar statements seem to be valid for NSVZ-like relations describing the renormalization of the gaugino mass \cite{Hisano:1997ua,Jack:1997pa,Avdeev:1997vx} in theories with softly broken supersymmetry \cite{Nartsev:2016nym,Nartsev:2016mvn}.

Knowing the prescription for constructing an NSVZ scheme we can significantly simplify the calculation of the $\beta$-function in higher loops. Really, the NSVZ equation relates the $\beta$-function to the anomalous dimension of the matter superfields in the previous orders. Therefore, say, for calculating the three-loop $\beta$-function it is sufficient to find the anomalous dimension of the matter superfields in the two-loop approximation. However, for this purpose it is necessary to use a renormalization scheme for which the NSVZ relation is valid, for instance, the HD+MSL prescription. For theories with simple gauge groups the three-loop $\beta$-function was calculated by this method in \cite{Kazantsev:2020kfl}. (Note that its part found earlier by direct calculations of three-loop supergraphs \cite{Shakhmanov:2017soc,Kazantsev:2018nbl} exactly coincided with the result obtained from the NSVZ equation. This confirms the correctness of the approach in such an approximation where the scheme dependence becomes essential.) For ${\cal N}=1$ SQED with $N_f$ flavors, the four-loop $\beta$-function was obtained from the NSVZ equation in \cite{Shirokov:2022jyd}. However, the NSVZ equations are also valid for ${\cal N}=1$ supersymmetric theories with multiple gauge couplings \cite{Ghilencea:1999cy}. For MSSM the equations equivalent to the NSVZ relations have first been written in \cite{Shifman:1996iy}. The standard form of these equations for MSSM and for the flipped $SU(5)$ model \cite{Barr:1981qv,Antoniadis:1987dx,Campbell:1987eb,Ellis:1988tx} can be found in \cite{Korneev:2021zdz}. There are strong evidences \cite{Korneev:2021zdz} that they are also satisfied in the HD+MSL scheme, exactly as for theories with simple gauge groups. For example, this is true for the NSVZ-like equation which gives the Adler $D$-function in ${\cal N}=1$ SQCD \cite{Shifman:2014cya,Shifman:2015doa} (see also \cite{Kataev:2017qvk}), which follows from the NSVZ equations for the theory with the gauge group $SU(N_c)\times U(1)$.

Because the NSVZ equations seem to be valid in the HD+MSL scheme, it is possible to use them for calculating quantum corrections even in theories with multiple gauge couplings. Then, making finite renormalizations one can obtain the $\beta$-functions for other renormalization prescriptions \cite{Vladimirov:1979ib,Vladimirov:1979my}. The results for RGFs in various subtraction schemes can be useful for investigating finiteness \cite{Parkes:1984dh,Parkes:1985hj,Parkes:1985hh,Kazakov:1986bs,Ermushev:1986cu,Lucchesi:1987he,Lucchesi:1987ef,Grisaru:1985tc,Jack:1996qq,Stepanyantz:2021dus}, the possibilities of the existence of fixed points \cite{Seiberg:1994pq,Ryttov:2012qu,Ryttov:2017khg,Bond:2022xvr}, or reduction of couplings \cite{Mondragon:2013aea,Heinemeyer:2014vxa,Heinemeyer:2017gsv,Heinemeyer:2018roq,Heinemeyer:2019vbc} in various ${\cal N}=1$ supersymmetric theories.

In this paper we will construct the three-loop MSSM $\beta$-functions for an arbitrary supersymmetric renormalization prescription. This implies that the theory is quantized in a manifestly ${\cal N}=1$ supersymmetric way with the help of ${\cal N}=1$ superspace \cite{Gates:1983nr,West:1990tg,Buchbinder:1998qv} and instead of renormalizing individual component fields we renormalize superfields as a whole. Equivalently, the renormalization constants for all components of a superfield appear to be the same. Evidently, the $\overline{\mbox{DR}}$ renormalization prescription is supersymmetric. Earlier in the $\overline{\mbox{DR}}$ scheme the two-loop MSSM anomalous dimensions of the chiral superfields and the two-loop MSSM Yukawa $\beta$-functions was found in \cite{Bjorkman:1985mi}. As we already mentioned, the three-loop MSSM gauge $\beta$-functions in the $\overline{\mbox{DR}}$ scheme was obtaned in \cite{Jack:2004ch}. As we will see below, for a certain renormalization prescription the results obtained in this paper correctly reproduce all these expressions, thus providing an independent confirmation of the complicated calculations made in \cite{Bjorkman:1985mi,Jack:2004ch}.

The paper is organized as follows. In Sect. \ref{Section_RGFs_With_Multiple_Couplings} we briefly describe ${\cal N}=1$ supersymmetric theories with multiple gauge couplings and discuss various properties of RGFs in these theories. In particular, we recall the form of the NSVZ equations in this case and present an explicit expression for the two-loop anomalous dimension (defined in terms of the bare couplings) of the chiral superfields for the case of using the higher covariant derivative regularization. The exact NSVZ $\beta$-functions for MSSM are written in Sect. \ref{Section_MSSM_NSVZ}. The two-loop anomalous dimensions for all chiral matter superfields in MSSM regularized by higher covariant derivatives are obtained in Sect. \ref{Section_MSSM_Anomalous_Dimensions}. We present both the anomalous dimensions defined in terms of the bare couplings and the ones defined in terms of the renormalized couplings for an arbitrary renormalization prescription. In Sect. \ref{Section_MSSM_Beta_Functions} the NSVZ equations are used for obtaining the three-loop MSSM $\beta$-functions defined in terms of the bare couplings. After that, we calculate the three-loop $\beta$-functions defined in terms of the renormalized couplings for an arbitrary renormalization prescription (supplementing the higher covariant derivative regularization). A subtraction scheme in which the results reproduce the ones in the $\overline{\mbox{DR}}$ scheme is constructed in Sect. \ref{Section_DR}. Finally, in Sect. \ref{Section_NSVZ_Class} we describe a class of the NSVZ schemes for MSSM and demonstrate that finite renormalizations relating various NSVZ schemes satisfy the general equations derived in \cite{Korneev:2021zdz}. The results for the two-loop anomalous dimensions of the matter superfields and the three-loop $\beta$-functions defined in terms of the bare couplings are presented in Appendices \ref{Appendix_Two_Loop_Gamma} and \ref{Appendix_Three_Loop_Beta}, respectively. For completeness, the expressions for RGFs in the $\overline{\mbox{DR}}$ scheme are listed in Appendix \ref{Appendix_RGFs_DR}.

\section{RGFs for theories with multiple gauge couplings}
\hspace*{\parindent}\label{Section_RGFs_With_Multiple_Couplings}

A part of the MSSM action which does not contain soft terms is an ${\cal N}=1$ supersymmetric Yang--Mills theory with 3 gauge couplings interacting with a certain set of chiral superfields. In general, a number of the gauge couplings is equal to the number of (simple or $U(1)$) factors in the gauge group

\begin{equation}\label{Gauge_Group_General}
G = G_1\times G_2 \times \ldots \times G_n.
\end{equation}

\noindent
The classical action of a renormalizable ${\cal N}=1$ supersymmetric gauge theory in the massless limit can be written in the form

\begin{eqnarray}\label{Classical_Action_General}
&& S = \sum\limits_{K=1}^n  \mbox{Re}\, \frac{1}{4}\int d^4x\, d^2\theta\, \left(W^a\right)^{A_K} \left(W_a\right)^{A_K} + \frac{1}{4} \int d^4x\, d^4\theta\, \phi^{*i} \big(e^{2V}\big)_{i}{}^{j} \phi_j\qquad\nonumber\\
&&\qquad\qquad\qquad\qquad\qquad\qquad\qquad\qquad\qquad\  +\Big(\frac{1}{6}\lambda_0^{ijk} \int d^4x\, d^2\theta\, \phi_i \phi_j \phi_k +\mbox{c.c.}\Big),\qquad
\end{eqnarray}

\noindent
where the subscript $K$ numerates the factors $G_K$ in the product (\ref{Gauge_Group_General}). The superfield formalism (see, e.g., \cite{Gates:1983nr,West:1990tg,Buchbinder:1998qv}) used here is very convenient because it makes ${\cal N}=1$ supersymmetry manifest. It is important that this property remains valid even at the quantum level.

In our notation the bare gauge and Yukawa couplings are denoted by $e_{0K}$ (or $\alpha_{0K}\equiv e_{0K}^2/4\pi$)\footnote{For the MSSM gauge coupling constant corresponding to the $U(1)$ subgroup below we will use a different definition.} and $\lambda_0^{ijk}$, respectively, and

\begin{equation}
V_{i}{}^j \equiv \sum\limits_K e_{0K} V^{A_K} \left(T^{A_K}\right)_i{}^j
\end{equation}

\noindent
is the gauge superfield with the strength $W_a$.

It is convenient to split chiral superfields $\phi$ into sets $\phi_{\mbox{\scriptsize a}}$ such that each of them either
transforms under certain irreducible representations $R_{\mbox{\scriptsize a} K}$ of the simple subgroups $G_K$ or has certain charges $q_{\mbox{\scriptsize a} K}$ with respect to $G_K = U(1)$. Then the index $i$ numerating all matter superfields $\phi$ can be presented as the set

\begin{equation}
i = \{\mbox{a};\, i_1, i_2, \ldots, i_n\} \equiv \{\mbox{a};\, i_{\mbox{\scriptsize a}}\}.
\end{equation}

\noindent
Note that the indices $i_1, i_2,\ldots, i_n$ are different for different $\mbox{a}$. The generators of the gauge group can be written as

\begin{equation}
\left(T^{A_K}\right)_i{}^j = \delta_{\mbox{\scriptsize a}}{}^{\mbox{\scriptsize b}}\cdot \delta_{i_1}{}^{j_1}\ldots \delta_{i_{K-1}}{}^{j_{K-1}} \left(T_{\mbox{\scriptsize a}}^{A_K}\right)_{i_K}{}^{j_K} \delta_{i_{K+1}}{}^{j_{K+1}}\ldots \delta_{i_n}{}^{j_n},
\end{equation}

\noindent
where $\left(T_{\mbox{\scriptsize a}}^{A_K}\right)_{i_K}{}^{j_K}$ are either the generators of $G_K$ in the representation $R_{\mbox{\scriptsize a} K}$ for simple subgroups $G_K$ or the charges $q_{\mbox{\scriptsize a} K}$ of the superfields $\phi_i \equiv \phi_{\mbox{\scriptsize a};\, i_1 i_2 \ldots i_n}$ for $G_K = U(1)$. They satisfy the commutation relations

\begin{equation}
[T^{A_K}, T^{B_K}] = i f^{A_K B_K C_K} T^{C_K};\qquad [T_{\mbox{\scriptsize a}}^{A_K}, T_{\mbox{\scriptsize a}}^{B_K}] = i f^{A_K B_K C_K} T_{\mbox{\scriptsize a}}^{C_K},
\end{equation}

\noindent
where $f^{A_K B_K C_K}$ are the structure constants of the subgroup $G_K$.

The renormalization constants for the matter superfields $\phi_{\mbox{\scriptsize a}}$ are introduced as $\phi_{\mbox{\scriptsize a}} = (\sqrt{Z})_{\mbox{\scriptsize a}}{}^{\mbox{\scriptsize b}} \phi_{\mbox{\scriptsize b}, R}$, where the subscript $R$ marks the renormalized superfields. The renormalization of each coupling constant $\alpha_K$ is described by the corresponding $\beta$-function, while the renormalization of the superfields $\phi_{\mbox{\scriptsize a}}$  is encoded in the corresponding anomalous dimension. In terms of the bare couplings these RGFs are defined by the equations

\begin{equation}\label{RGFs_Bare_Definition}
\beta_K(\alpha_0,\lambda_0) \equiv \left.\frac{d\alpha_{0K}}{d\ln\Lambda}\right|_{\alpha,\lambda = \mbox{\scriptsize const}};\qquad
\gamma_{\mbox{\scriptsize a}}{}^{\mbox{\scriptsize b}}(\alpha_0,\lambda_0) \equiv -\frac{d\ln Z_{\mbox{\scriptsize a}}{}^{\mbox{\scriptsize b}}}{d\ln\Lambda}\bigg|_{\alpha,\lambda = \mbox{\scriptsize const}},
\end{equation}

\noindent
where the derivatives are taken with respect to the dimensionful regularization parameter $\Lambda$ at fixed values of the renormalized couplings. According to \cite{Kataev:2013eta}, RGFs defined in terms of the bare couplings should be distinguished from the standard ones, which are defined in terms of the renormalized couplings by the equations

\begin{equation}\label{RGFs_Standard_Definition}
\widetilde\beta_K(\alpha,\lambda) \equiv \left.\frac{d\alpha_K}{d\ln\mu}\right|_{\alpha_0,\lambda_0 = \mbox{\scriptsize const}};\qquad
\widetilde\gamma_{\mbox{\scriptsize a}}{}^{\mbox{\scriptsize b}}(\alpha,\lambda) \equiv \frac{d\ln Z_{\mbox{\scriptsize a}}{}^{\mbox{\scriptsize b}}}{d\ln\mu}\bigg|_{\alpha_0,\lambda_0 = \mbox{\scriptsize const}}.
\end{equation}

\noindent
In this case the differentiations are made with respect to the renormalization point $\mu$ at fixed values of bare couplings.

Note that for $\mbox{a}\ne \mbox{b}$ the anomalous dimension $\widetilde\gamma_{\mbox{\scriptsize a}}{}^{\mbox{\scriptsize b}}(\alpha,\lambda)$ (or $\gamma_{\mbox{\scriptsize a}}{}^{\mbox{\scriptsize b}}(\alpha_0,\lambda_0)$) does not vanish only if the representations $R_{\mbox{\scriptsize a}K}$ and $R_{\mbox{\scriptsize b}K}$ (or the corresponding charges for $U(1)$ subgroups) coincide for all $K=1,\ldots,n$. Even in this case (corresponding to the existence of some generations of the superfields) the anomalous dimension matrix can be diagonalized by a certain rotation in the generation space. However, the standard choice of variables in MSSM corresponds to the diagonal mass matrix for the charged leptons. In this case the anomalous dimension matrix is not in general diagonal. That is why below we will write the anomalous dimension in the matrix form in contrast with Ref. \cite{Korneev:2021zdz}, where this matrix was assumed to be diagonalized.

As we mentioned above, there are strong evidences (see, e.g., \cite{Korneev:2021zdz}) that in the case of using the higher covariant derivative regularization RGFs defined in terms of the bare couplings satisfy the NSVZ equations

\begin{equation}\label{NSVZ_Multicharge_General}
\frac{\beta_K(\alpha_0,\lambda_0)}{\alpha_{0K}^2} = - \frac{1}{2\pi(1-C_{2}(G_K) \alpha_{0K}/2\pi)} \Big[\, 3 C_2(G_K) - \sum\limits_{\mbox{\scriptsize a}} \bm{T}_{\mbox{\scriptsize a}K}\Big(1-\gamma_{\mbox{\scriptsize a}}{}^{\mbox{\scriptsize a}}(\alpha_0,\lambda_0)\Big) \Big]
\end{equation}

\noindent
in all orders of the perturbation theory independently of a renormalization prescription. (Note that RGFs (\ref{RGFs_Bare_Definition}) depend on a regularization, but do not depend on a renormalization prescription for a fixed regularization, see \cite{Kataev:2013eta} for details.) In Eq. (\ref{NSVZ_Multicharge_General}) we use the notation

\begin{eqnarray}\label{Casimirs_Definitions}
&& C_2(G_K) \delta^{A_K B_K} = f^{A_K C_K D_K} f^{B_K C_K D_K};\qquad  T_K(R_{\mbox{\scriptsize a} K})\, \delta^{{A_K}{B_K}} = (T_{\mbox{\scriptsize a}}^{A_K} T_{\mbox{\scriptsize a}}^{B_K})_{i_K}{}^{i_K};\vphantom{\frac{1}{p}}\nonumber\\
&& \bm{T}_{\mbox{\scriptsize a}K} = \left\{
\begin{array}{l}
{\displaystyle \delta_{i_1}{}^{i_1}\ldots \delta_{i_{K-1}}{}^{i_{K-1}} T_K(R_{\mbox{\scriptsize a} K})\, \delta_{i_{K+1}}{}^{i_{K+1}}\ldots \delta_{i_n}{}^{i_n}\qquad \mbox{for a simple}\ G_K;}\\
\\
{\displaystyle \delta_{i_1}{}^{i_1}\ldots \delta_{i_{K-1}}{}^{i_{K-1}}\, q_{\mbox{\scriptsize a} K}^2\, \delta_{i_{K+1}}{}^{i_{K+1}}\ldots \delta_{i_n}{}^{i_n}\qquad\qquad\ \mbox{for}\ G_K = U(1).}
\end{array}
\right.
\end{eqnarray}

It is very convenient to use Eq. (\ref{NSVZ_Multicharge_General}) for calculating the $\beta$-functions because it relates them to the anomalous dimensions of the matter superfields {\it in the previous orders}. Therefore, for obtaining, e.g., the three-loop $\beta$-functions it is sufficient to calculate the anomalous dimensions in the two-loop approximation. However, in this case we have to use the higher covariant derivative regularization. To introduce it, we should make the following steps.

First, we make the background-quantum splitting of the gauge superfield $e^{2V} \to e^{2{\cal F}(V)} e^{2\bm{V}}$, where $\bm{V}$ is the background gauge superfield, and the function ${\cal F}(V)$ is needed because the quantum gauge superfield $V$ is renormalized in a nonlinear way \cite{Piguet:1981fb,Piguet:1981hh,Tyutin:1983rg}. The explicit form of this function in the lowest nontrivial order can be found in \cite{Juer:1982fb,Juer:1982mp}. In particular, in this approximation it contains a term cubic in the gauge superfield with a new bare parameter $y_0$, which is similar to the gauge parameter $\xi_0$. In general, there are an infinite number of nonlinear terms and similar parameters in the function ${\cal F}(V)$. If the renormalization of these parameters is not taken into account, then standard RGFs cannot be written as functions of the renormalized couplings (and other parameters) only and will explicitly depend on $\ln\Lambda/\mu$ \cite{Kazantsev:2018kjx}. By other words, the renormalization group equations will not be satisfied without the nonlinear terms in the function ${\cal F}(V)$. Next, we modify the action by adding some terms with higher covariant derivatives,

\begin{eqnarray}\label{Regularized_Action_With_Multiple_Couplings}
&&\hspace*{-7mm} S\to S_{\mbox{\scriptsize reg}} = \sum\limits_{K=1}^n\mbox{Re}\, \frac{1}{4} \int d^4x\, d^2\theta \left(W^a\right)^{A_K} \Big[\Big(e^{-2\bm{V}} e^{-2{\cal F}(V)} R\Big(-\frac{\bar\nabla^2 \nabla^2}{16\Lambda^2}\Big) e^{2{\cal F}(V)} e^{2\bm{V}}\Big)_{Adj} W_{a}\Big]^{A_K}\nonumber\\
&&\hspace*{-7mm} + \frac{1}{4} \int d^4x\, d^4\theta\, \phi^{*i} \Big[F\Big(-\frac{\bar\nabla^2 \nabla^2}{16\Lambda^2}\Big) e^{2{\cal F}(V)} e^{2\bm{V}}\Big]_i{}^j \phi_j +\Big(\frac{1}{6} \lambda_0^{ijk} \int d^4x\, d^2\theta\, \phi_{i} \phi_{j} \phi_{k} +\mbox{c.c.}\Big),
\end{eqnarray}

\noindent
where $R(x)$ and $F(x)$ are the regulator functions which rapidly grow at infinity and are equal to 1 at $x=0$. Note that, for simplicity, we use the same regulator functions for all subgroups $G_K$ in the product (\ref{Gauge_Group_General}). The covariant derivatives entering Eq. (\ref{Regularized_Action_With_Multiple_Couplings}) are defined by the equations

\begin{equation}
\nabla_a = D_a;\qquad \bar\nabla_{\dot a} = e^{2{\cal F}(V)} e^{2\bm{V}} \bar D_{\dot a} e^{-2\bm{V}} e^{-2 {\cal F}(V)}.
\end{equation}

Next, we add the gauge fixing term $S_{\mbox{\scriptsize gf}}$, the action for the Faddeev--Popov ghosts $S_{\mbox{\scriptsize FP}}$, and the action for the Nielsen--Kallosh ghosts $S_{\mbox{\scriptsize NK}}$. Finally, it is necessary to insert into the generating functional the Pauli--Villars determinants which remove residual one-loop divergences and subdivergences. The resulting expression for the generating functional can be written as

\begin{equation}
Z[\mbox{sources}] = \int D\mu\,\prod\limits_K \mbox{Det}^{c_K}(PV,M_K)\, \exp\Big\{i\Big(S_{\mbox{\scriptsize reg}} + S_{\mbox{\scriptsize gf}} + S_{\mbox{\scriptsize FP}} + S_{\mbox{\scriptsize NK}} + S_{\varphi} + S_{\mbox{\scriptsize sources}}\Big)\Big\}.
\end{equation}

\noindent
The Pauli--Villars superfields $\varphi_{1,K}$, $\varphi_{2,K}$, and $\varphi_{3,K}$ with the action $S_{\varphi}$ are introduced for all $K$ corresponding to simple subgroups of the gauge group $G$. These superfields lie in the adjoint representation of the subgroup $G_K$ and are neutral with respect to the other subgroups. They have the masses $M_{\varphi, K} = a_{\varphi, K}\Lambda$ and cancel one-loop divergences coming from a loop of gauge and ghost superfields. The Pauli--Villars superfields inside the determinants $\mbox{Det}(PV,M_K)$ have the masses $M_K = a_K \Lambda$ and remove the one-loop divergences produced by matter superfields. The detailed description of the Pauli--Villars determinants and expressions for the coefficients $c_K$ can be found in \cite{Aleshin:2016yvj,Kazantsev:2017fdc} for theories with a single gauge coupling and in \cite{Korneev:2021zdz} for theories with multiple gauge couplings.

For theories with a single gauge coupling constant regularized by higher covariant derivatives the two-loop anomalous dimension of the matter superfields defined in terms of the bare couplings has been calculated in \cite{Kazantsev:2020kfl}. The result is given by the expression

\begin{eqnarray}\label{Two_Loop_Gamma_Bare_Single_Gauge_Coupling}
&&\hspace*{-7mm} \gamma_i{}^j(\alpha_0,\lambda_0) = - \frac{d\ln Z_i{}^j}{d\ln\Lambda}\bigg|_{\alpha,\lambda=\mbox{\scriptsize const}} = - \frac{\alpha_0}{\pi}C(R)_i{}^j+\frac{1}{4\pi^2}\lambda^*_{0imn}\lambda_0^{jmn}  + \frac{\alpha_0^2}{2\pi^2} \left[C(R)^2\right]_i{}^j -\frac{3\alpha_0^2}{2\pi^2}\, C_2 \nonumber\\
&&\hspace*{-7mm} \times C(R)_i{}^j\Big(\ln a_{\varphi}+1+\frac{A}{2}\Big) +\frac{\alpha_0^2}{2\pi^2}\, T(R) C(R)_i{}^j\Big(\ln a + 1 + \frac{A}{2}\Big) - \frac{\alpha_0}{8\pi^3} \lambda^*_{0lmn}\lambda^{jmn}_0 C(R)_i{}^l (1-B \nonumber\\
&&\hspace*{-7mm} +A) + \frac{\alpha_0}{4\pi^3}\lambda^*_{0imn}\lambda_0^{jml}C(R)_l{}^n(1-A+B) -\frac{1}{16\pi^4}\lambda^*_{0iac}\lambda_0^{jab}\lambda^*_{0bde}\lambda_0^{cde}
+ O\Big(\alpha_0^3,\alpha_0^2\lambda_0^2,\alpha_0\lambda_0^4,\lambda_0^6\Big),\nonumber\\
\end{eqnarray}

\noindent
where $C(R)_i{}^j \equiv (T^A T^A)_i{}^j$, $\mbox{tr}(T^A T^B) = \delta^{AB} T(R)$, and $T(Adj) \equiv C_2$. The parameters $A$ and $B$ are related to the regulator functions $R(x)$ and $F(x)$ by the equations

\begin{equation}\label{AB_Definition}
A\equiv \int\limits_0^\infty dx \ln x\, \frac{d}{dx}\frac{1}{R(x)};\qquad B \equiv \int\limits_0^\infty dx \ln x\, \frac{d}{dx}\frac{1}{F^2(x)},
\end{equation}

\noindent
and the parameters $a$ and $a_\varphi$ are the (above introduced) ratios of the Pauli--Villars masses to the regularization parameter $\Lambda$,

\begin{equation}
a \equiv \frac{M}{\Lambda};\qquad a_\varphi \equiv \frac{M_\varphi}{\Lambda}.
\end{equation}

The expression (\ref{Two_Loop_Gamma_Bare_Single_Gauge_Coupling}) can be generalized to the case of theories with multiple gauge couplings. First, we rewrite the Yukawa part of the action (\ref{Classical_Action_General}) in terms of the superfields $\phi_{\mbox{\scriptsize a}}$,

\begin{equation}
\frac{1}{6} \lambda_0^{ijk} \int d^4x\, d^2\theta\, \phi_i\phi_j\phi_k\ \to\ \frac{1}{6}\sum\limits_{\mbox{\scriptsize abc}} \lambda_0^{i_{\mbox{\tiny a}} j_{\mbox{\tiny b}} k_{\mbox{\tiny c}}} \int d^4x\, d^2\theta\, \phi_{a;i_{\mbox{\tiny a}}} \phi_{b;j_{\mbox{\tiny b}}} \phi_{c;k_{\mbox{\tiny c}}}.
\end{equation}

\noindent
Next, it is necessary to reanalyse contributions of various supergraphs to the two-loop anomalous dimension. Namely, we should take into account that the expression for a supergraph can contain various gauge coupling constants and specify the gauge group factors. The result obtained after this can be presented in the form

\begin{eqnarray}\label{Gamma_Multicharge}
&&\hspace*{-5mm} \gamma_{\mbox{\scriptsize a}}{}^{\mbox{\scriptsize b}}(\alpha_0,\lambda_0) = - \sum\limits_K \frac{\alpha_{0K}}{\pi} C(R_{\mbox{\scriptsize a} K})\, \delta_{\mbox{\scriptsize a}}{}^{\mbox{\scriptsize b}} + \frac{1}{4\pi^2} \big(\lambda_0^*\lambda_0\big)_{\mbox{\scriptsize a}}{}^{\mbox{\scriptsize b}}
+ \sum\limits_{KL}\frac{\alpha_{0K} \alpha_{0L}}{2\pi^2} C(R_{\mbox{\scriptsize a} K}) C(R_{\mbox{\scriptsize a} L})\, \delta_{\mbox{\scriptsize a}}{}^{\mbox{\scriptsize b}}
\nonumber\\
&&\hspace*{-5mm} - \sum\limits_K \frac{3\alpha_{0K}^2}{2\pi^2}\, C_2(G_K) C(R_{\mbox{\scriptsize a} K}) \Big(\ln a_{\varphi, K}+1+\frac{A}{2}\Big)\, \delta_{\mbox{\scriptsize a}}{}^{\mbox{\scriptsize b}} + \sum\limits_K \frac{\alpha_{0K}^2}{2\pi^2} C(R_{\mbox{\scriptsize a} K}) \sum\limits_{\mbox{\scriptsize c}}\bm{T}_{\mbox{\scriptsize c}K} \Big(\ln a_K \nonumber\\
&&\hspace*{-5mm} + 1 + \frac{A}{2}\Big)\, \delta_{\mbox{\scriptsize a}}{}^{\mbox{\scriptsize b}}
- \sum\limits_K \frac{\alpha_{0K}}{8\pi^3}  (\lambda_0^*\lambda_0)_{\mbox{\scriptsize a}}{}^{\mbox{\scriptsize b}} C(R_{\mbox{\scriptsize a} K}) (1-B+A) + \sum\limits_K \frac{\alpha_{0K}}{4\pi^3} \big(\lambda_0^* C_K\lambda_0\big)_{\mbox{\scriptsize a}}{}^{\mbox{\scriptsize b}} (1-A+B)
\nonumber\\
&&\hspace*{-5mm} -\frac{1}{16\pi^4}\big(\lambda_0^* [\lambda_0^*\lambda_0]\lambda_0\big)_{\mbox{\scriptsize a}}{}^{\mbox{\scriptsize b}} + O(\alpha_0^3,\alpha_0^2\lambda_0^2,\alpha_0\lambda_0^4,\lambda_0^6), \vphantom{\frac{1}{2}}
\end{eqnarray}

\noindent
where we use the notations

\begin{eqnarray}\label{C(R)_Definition}
&& (T_{\mbox{\scriptsize a}}^{A_K} T_{\mbox{\scriptsize a}}^{A_K})_{i_K}{}^{j_K} = C(R_{\mbox{\scriptsize a} K})\, \delta_{i_K}{}^{j_K};\vphantom{\sum\limits_{\mbox{\scriptsize bc}}}\\
\label{Lambda0_Definition}
&& \big(\lambda_0^* \lambda_0\big)_{\mbox{\scriptsize a}}{}^{\mbox{\scriptsize b}}\, \delta_{i_{\mbox{\tiny a}}}{}^{j_{\mbox{\tiny b}}} = \sum\limits_{\mbox{\scriptsize cd}} \lambda^*_{0\, i_{\mbox{\tiny a}} m_{\mbox{\tiny c}}  n_{\mbox{\tiny d}}} \lambda_0^{j_{\mbox{\tiny b}} m_{\mbox{\tiny c}}  n_{\mbox{\tiny d}}};\\
\label{Lambda1_Definition}
&& \big(\lambda_0^* C_K\lambda_0\big)_{\mbox{\scriptsize a}}{}^{\mbox{\scriptsize b}}\, \delta_{i_{\mbox{\tiny a}}}{}^{j_{\mbox{\tiny b}}}  = \sum\limits_{\mbox{\scriptsize cd}}
\lambda^*_{0\, i_{\mbox{\tiny a}} m_{\mbox{\tiny c}}  n_{\mbox{\tiny d}}} C(R_{\mbox{\scriptsize d} K}) \lambda_0^{j_{\mbox{\tiny b}} m_{\mbox{\tiny c}}  n_{\mbox{\tiny d}}};\\
\label{Lambda2_Definition}
&& \big(\lambda_0^* [\lambda_0^*\lambda_0]\lambda_0\big)_{\mbox{\scriptsize a}}{}^{\mbox{\scriptsize b}}\, \delta_{i_{\mbox{\tiny a}}}{}^{j_{\mbox{\tiny b}}} = \sum\limits_{\mbox{\scriptsize cdefg}}
\lambda^*_{0i_{\mbox{\tiny a}} k_{\tiny e} l_{\mbox{\tiny f}}}
\lambda_0^{j_{\mbox{\tiny b}} k_{\mbox{\tiny e}} p_{\mbox{\tiny g}}}
\lambda^*_{0 p_{\mbox{\tiny g}} m_{\mbox{\tiny c}} n_{\mbox{\tiny d}}}
\lambda_0^{l_{\mbox{\tiny f}} m_{\mbox{\tiny c}} n_{\mbox{\tiny d}}}.
\end{eqnarray}

\noindent
Note that for $\mbox{a}\ne \mbox{b}$ in Eqs. (\ref{Lambda0_Definition}) --- (\ref{Lambda2_Definition}) the indices $i_{\mbox{\tiny a}}$ and $j_{\mbox{\tiny b}}$ can coincide only if the superfields $\phi_{\mbox{\scriptsize a}}$ and $\phi_{\mbox{\scriptsize b}}$ have the same quantum numbers. (Certainly, in MSSM this corresponds to the same superfields of different generations.)

Below we will use Eq. (\ref{Gamma_Multicharge}) as a starting point for calculating the two-loop anomalous dimensions of the MSSM chiral matter superfields.

\section{The exact $\beta$-functions in MSSM}
\hspace*{\parindent}\label{Section_MSSM_NSVZ}

MSSM (see, e.g., \cite{Mohapatra:1986uf}) is a softly broken ${\cal N}=1$ supersymmetric theory with the gauge group

\begin{equation}\label{MSSM_Gauge_Group}
G = SU(3) \times SU(2) \times U(1)_Y
\end{equation}

\noindent
and chiral matter superfields listed in Table \ref{Table_MSSM_Chiral_Superfields}, where we also present their quantum numbers with respect to the gauge group. The chiral superfields include three generations of quarks and leptons, Higgs fields, and their superpartners as components. (Note that we consider a model without right neutrinos.) The chiral superfields in our notation are denoted by capital letters, and the subscripts $1,2,3$ numerate generations.

\begin{table}[h]
\begin{center}
\begin{tabular}{|c||c|c|c|c|c||c|c|}
\hline
$\mbox{superfield}\vphantom{\Big(}$ & $Q_1,Q_2,Q_3$ & $U_1, U_2, U_3$ & $D_1, D_2, D_3$ & $L_1, L_2, L_3$ & $E_1, E_2, E_3$ &\ \ \ $H_u$\ \ \ &\ \ \ $H_d$\ \ \  \\
\hline
$SU(3)\vphantom{\Big(}$ & $\bar 3$ & 3 & 3 & 1 & 1 & 1 & 1 \\
\hline
$SU(2)\vphantom{\Big(}$ & 2 & 1 & 1 & 2 & 1 & 2 & 2 \\
\hline
$U(1)_Y\vphantom{\Big(}$ & $-1/6$ & $2/3$ & $-1/3$ & $1/2$ & $-1$ & $-1/2$ & $1/2$\\
\hline
\end{tabular}
\end{center}
\caption{Chiral matter superfields and their quantum numbers (representations for $SU(3)$ and $SU(2)$, and the hypercharge $Y$ for $U(1)$) in MSSM.}\label{Table_MSSM_Chiral_Superfields}
\end{table}

Because the gauge group (\ref{MSSM_Gauge_Group}) is a product of three factors, the theory contains three gauge coupling constants

\begin{equation}\label{Couplings_Definitions}
\alpha_3 = \frac{e_3^2}{4\pi};\qquad \alpha_2 = \frac{e_2^2}{4\pi};\qquad  \alpha_1 = \frac{5}{3}\cdot \frac{e_1^2}{4\pi}.
\end{equation}

\noindent
Note that it is reasonable to include the factor $5/3$ into the definition of $\alpha_1$ because in this case the gauge coupling unification condition takes the most convenient form $\alpha_1 = \alpha_2 = \alpha_3$.

A part of the MSSM action

\begin{equation}
\Delta S = \frac{1}{2} \int d^4x\, d^2\theta\, W + \mbox{c.c.}
\end{equation}

\noindent
contains the superpotential

\begin{eqnarray}\label{MSSM_Superpotential}
&&\hspace*{-11mm} W = \left(Y_{0U}\right)_{IJ}
\left(\widetilde U\ \widetilde D \right)^{a}_I
\left(
\begin{array}{cc}
0 & 1\\
-1 & 0
\end{array}
\right)
\left(
\begin{array}{c}
H_{u1}\\ H_{u2}
\end{array}
\right) U_{aJ}
+ \left(Y_{0D}\right)_{IJ}
\left(\widetilde U\ \widetilde D \right)^{a}_I
\left(
\begin{array}{cc}
0 & 1\\
-1 & 0
\end{array}
\right)
\left(
\begin{array}{c}
H_{d1}\\ H_{d2}
\end{array}
\right)
\nonumber\\
&&\hspace*{-11mm} \times D_{aJ} + \left(Y_{0E}\right)_{IJ} \left(\widetilde N\ \widetilde E \right)_{I}
\left(
\begin{array}{cc}
0 & 1\\
-1 & 0
\end{array}
\right)
\left(
\begin{array}{c}
H_{d1}\\ H_{d2}
\end{array}
\right) E_J
+ \bm{\mu}_0 \left(H_{u1}\ H_{u2} \right)
\left(
\begin{array}{cc}
0 & 1\\
-1 & 0
\end{array}
\right)
\left(
\begin{array}{c}
H_{d1}\\ H_{d2}
\end{array}
\right).
\end{eqnarray}

\noindent
Note that in this expression we presented the superfields which include left quarks and leptons as

\begin{equation}
Q = \left(\begin{array}{c}\widetilde U\\ \widetilde D\end{array}\right);\qquad L = \left(\begin{array}{c}\widetilde N\\ \widetilde E\end{array}\right),
\end{equation}

\noindent
respectively, and wrote the superfields $H_u$ and $H_d$ (in the fundamental representation of $SU(2)$) as the two-component columns. Also in Eq. (\ref{MSSM_Superpotential}) $Y_{0U}$, $Y_{0D}$, and $Y_{0E}$ are the (dimensionless) Yukawa matrices with the indices $I$ and $J$, which numerate generations and take values from 1 to 3. The parameter $\bm{\mu}_0$ has the dimension of mass.

According to \cite{Shifman:1996iy,Ghilencea:1997mu,Ghilencea:1999cy} it is possible to construct the exact NSVZ expressions for the $\beta$-functions for all couplings, which can be written in the form \cite{Korneev:2021zdz}

\begin{eqnarray}\label{MSSM_Exact_Beta3}
&&\hspace*{-5mm} \frac{\beta_3(\alpha_0,Y_0)}{\alpha_{03}^2} = - \frac{1}{2\pi(1 - 3\alpha_{03}/2\pi)} \bigg[3 + \mbox{tr}\Big(\gamma_{Q}(\alpha_0,Y_0) + \frac{1}{2} \gamma_{U}(\alpha_0,Y_0) + \frac{1}{2} \gamma_{D}(\alpha_0,Y_0)\Big)\bigg];\\
\label{MSSM_Exact_Beta2}
&&\hspace*{-5mm} \frac{\beta_2(\alpha_0,Y_0)}{\alpha_{02}^2} = - \frac{1}{2\pi(1 - \alpha_{02}/\pi)} \bigg[-1 + \mbox{tr}\Big(\frac{3}{2} \gamma_{Q}(\alpha_0,Y_0) + \frac{1}{2} \gamma_{L}(\alpha_0,Y_0)\Big) + \frac{1}{2} \gamma_{H_u}(\alpha_0,Y_0)\nonumber\\
&& + \frac{1}{2} \gamma_{H_d}(\alpha_0,Y_0)\bigg];\\
\label{MSSM_Exact_Beta1}
&&\hspace*{-5mm} \frac{\beta_1(\alpha_0,Y_0)}{\alpha_{01}^2} = - \frac{3}{5} \cdot \frac{1}{2\pi}\bigg[-11 + \mbox{tr}\Big(\frac{1}{6} \gamma_{Q}(\alpha_0,Y_0) + \frac{4}{3} \gamma_{U}(\alpha_0,Y_0) + \frac{1}{3} \gamma_{D}(\alpha_0,Y_0) + \frac{1}{2} \gamma_{L}(\alpha_0,Y_0)  \nonumber\\
&&\hspace*{-5mm} + \gamma_{E}(\alpha_0,Y_0)\Big) + \frac{1}{2} \gamma_{H_u}(\alpha_0,Y_0) + \frac{1}{2} \gamma_{H_d}(\alpha_0,Y_0)\bigg],
\end{eqnarray}

\noindent
where the traces are taken over the indices numerating generations. These NSVZ equations are written for RGFs defined in terms of the bare couplings. There are strong evidences that with the higher covariant derivative regularization they are valid for an arbitrary renormalization prescription. (Actually, both sides of these equation do not depend on a renormalization prescription for a fixed version of the higher covariant derivative regularization.) The detailed description of the higher covariant derivative regularization used in this paper for MSSM can be found in \cite{Korneev:2021zdz}.

\section{Two-loop anomalous dimensions for various MSSM chiral superfields}
\hspace*{\parindent}\label{Section_MSSM_Anomalous_Dimensions}

From Eq. (\ref{Gamma_Multicharge}) it is possible to construct two-loop anomalous dimensions for all chiral superfields in MSSM. The bare Yukawa couplings $\lambda_0^{ijk}$ can be expressed in terms of the matrices $Y_{0U}$, $Y_{0D}$, and $Y_{0E}$. The constants $C_2(G_K)$ and $C(R_{\mbox{\scriptsize a}K})$ can be found using the equations

\begin{eqnarray}
&& C_2(SU(N))=N; \qquad\qquad\qquad\quad C_2(U(1)) = 0;\qquad\nonumber\\
&& C\Big(\mbox{fund.}\, SU(N)\Big) = \frac{N^2-1}{2N};\qquad\, C_{\mbox{\scriptsize a}\,U(1)} = q_{\mbox{\scriptsize a}}^2.\qquad
\end{eqnarray}

\noindent
Values of the constants $\bm{T}_{\mbox{\scriptsize a} K}$ and $(\lambda^*\lambda)_{\mbox{\scriptsize a}}{}^{\mbox{\scriptsize b}}$ are listed in Tables \ref{Table_MSSM_TaK} and \ref{Table_MSSM_LambdaLambda}, respectively. (Certainly, for calculating $(\lambda_0^*\lambda_0)_{\mbox{\scriptsize a}}{}^{\mbox{\scriptsize b}}$ one should replace $Y$ by $Y_0$.) In our notation $M^T$ denotes the transpose of a matrix $M$ (with respect to the indices numerating generations).

The expressions for all anomalous dimensions of the MSSM matter superfields defined in terms of the bare couplings constructed with the help of Eq. (\ref{Gamma_Multicharge}) are rather large. That is why we present them in Appendix \ref{Appendix_Two_Loop_Gamma}. Note that we do not diagonalize them by special rotations in the generation space, so that for the quark and lepton superfields they are given by nondiagonal $3\times 3$ matrices.

\vspace{5mm}
\begin{table}[h]
\begin{center}
\begin{tabular}{|c||c|c|c|c|c||c|c|}
\hline
$\mbox{superfield}\vphantom{\Big(}$ & $Q_1,Q_2,Q_3$ & $U_1, U_2, U_3$ & $D_1, D_2, D_3$ & $L_1, L_2, L_3$ & $E_1, E_2, E_3$ &\ \ \ $H_u$\ \ \ &\ \ \ $H_d$\ \ \ \\
\hline
$SU(3)\vphantom{\Big(}$ & 1 & $1/2$ & $1/2$ & 0 & 0 & 0 & 0 \\
\hline
$SU(2)\vphantom{\Big(}$ & $3/2$ & 0 & 0 & $1/2$ & 0 & $1/2$ & $1/2$ \\
\hline
$U(1)_Y\vphantom{\Big(}$ & 1/6 & 4/3 & 1/3 & 1/2  & 1 & $1/2$ & $1/2$\\
\hline
\end{tabular}
\end{center}
\caption{$\bm{T}_{\mbox{\scriptsize a} K}$ for various MSSM superfields. The superfields are numerated by the subscript $\mbox{a}$, and $K = SU(3),\ SU(2),\, U(1)$.}\label{Table_MSSM_TaK}
\end{table}

\vspace{5mm}
\begin{table}[h]
\begin{center}
\begin{tabular}{|c||c|c|c|c|}
\hline
$\mbox{superfield}\vphantom{\Big(}$ & $Q_1,Q_2,Q_3$ & \quad $U_1, U_2, U_3 \quad $ & $D_1, D_2, D_3$ &\quad $L_1, L_2, L_3 \quad$  \\
\hline
$\lambda^*\lambda\vphantom{\Bigg(}$ & ${\displaystyle \frac{1}{2}}\left(Y_U Y_U^+ + Y_D Y_D^+\right)^T$ & $Y_U^+ Y_U$ & $Y_D^+ Y_D$ & ${\displaystyle \frac{1}{2}} \left(Y_E Y_E^+\right)^T$  \\
\hline
\hline
$\mbox{superfield}\vphantom{\Big(}$ & $E_1, E_2, E_3$ &\ \ \ $H_u$\ \ \ &\ \ \ $H_d$\ \ \ & \\
\hline
$\lambda^*\lambda\vphantom{\Bigg(}$ & $Y_E^+ Y_E$ & ${\displaystyle \frac{3}{2}}\mbox{tr}\left(Y_U^+ Y_U\right)$ & ${\displaystyle \frac{1}{2}} \mbox{tr}\left(3 Y_D^+ Y_D + Y_E^+ Y_E\right)$ & \\
\hline
\end{tabular}
\end{center}
\caption{Values of $(\lambda^*\lambda)_{\mbox{\scriptsize a}}{}^{\mbox{\scriptsize b}}$ for various MSSM superfields. The nondiagonal elements of this matrix can be nontrivial only if $\mbox{a}$ and $\mbox{b}$ correspond to the same superfields of different generations.}\label{Table_MSSM_LambdaLambda}
\end{table}

However, standard RGFs are defined in terms of the renormalized couplings. To obtain these RGFs, we should first integrate the renormalization group equations (\ref{RGFs_Bare_Definition}) and find the renormalization constants for the gauge couplings and for the chiral matter superfields. The renormalization constants constructed in this way should be substituted into Eq. (\ref{RGFs_Standard_Definition}). Note that they depend on some finite constants which specify a renormalization prescription in the considered approximation. In particular, in the lowest approximations the relations between the bare and renormalized gauge coupling constants can be written in the form

\begin{eqnarray}\label{Alpha3}
&&\hspace*{-5mm} \frac{1}{\alpha_{03}} =\frac{1}{\alpha_3} + \frac{1}{2\pi}\bigg[3\Big(\ln\frac{\Lambda}{\mu} + b_{1,3}\Big) -\frac{11\alpha_1}{20\pi}\Big(\ln\frac{\Lambda}{\mu} + b_{2,31}\Big) -\frac{9\alpha_2}{4\pi}\Big(\ln\frac{\Lambda}{\mu} + b_{2,32}\Big) - \frac{7\alpha_3}{2\pi}\Big(\ln\frac{\Lambda}{\mu}\nonumber\\
&&\hspace*{-5mm} + b_{2,33}\Big) +\frac{1}{4\pi^2}\mbox{tr}\Big(Y_U^+ Y_U\Big)\Big(\ln\frac{\Lambda}{\mu} + b_{2,3U}\Big) + \frac{1}{4\pi^2}\mbox{tr}\Big(Y_D^+ Y_D\Big)\Big(\ln\frac{\Lambda}{\mu} + b_{2,3D}\Big)\bigg] + O(\alpha^2,\alpha Y^2, Y^4);\nonumber\\
&&\vphantom{1}\\
\label{Alpha2}
&&\hspace*{-5mm} \frac{1}{\alpha_{02}} =\frac{1}{\alpha_2} + \frac{1}{2\pi}\bigg[- \Big(\ln\frac{\Lambda}{\mu} + b_{1,2}\Big) -\frac{9\alpha_1}{20\pi}\Big(\ln\frac{\Lambda}{\mu} + b_{2,21}\Big) -\frac{25\alpha_2}{4\pi}\Big(\ln\frac{\Lambda}{\mu} + b_{2,22}\Big) - \frac{6\alpha_3}{\pi}\Big(\ln\frac{\Lambda}{\mu}\nonumber\\
&&\hspace*{-5mm} + b_{2,23}\Big) +\frac{3}{8\pi^2}\mbox{tr}\Big(Y_U^+ Y_U\Big)\Big(\ln\frac{\Lambda}{\mu} + b_{2,2U}\Big) + \frac{3}{8\pi^2}\mbox{tr}\Big(Y_D^+ Y_D\Big)\Big(\ln\frac{\Lambda}{\mu} + b_{2,2D}\Big)
+ \frac{1}{8\pi^2}\mbox{tr}\Big(Y_E^+ Y_E\Big)\nonumber\\
&&\hspace*{-5mm} \times \Big(\ln\frac{\Lambda}{\mu} + b_{2,2E}\Big)\bigg] + O(\alpha^2,\alpha Y^2, Y^4);\\
\label{Alpha1}
&&\hspace*{-5mm} \frac{1}{\alpha_{01}} =\frac{1}{\alpha_1} + \frac{1}{2\pi}\cdot \frac{3}{5}\bigg[-11\Big(\ln\frac{\Lambda}{\mu} + b_{1,1}\Big) -\frac{199\alpha_1}{60\pi}\Big(\ln\frac{\Lambda}{\mu} + b_{2,11}\Big) -\frac{9\alpha_2}{4\pi}\Big(\ln\frac{\Lambda}{\mu} + b_{2,12}\Big) - \frac{22\alpha_3}{3\pi}\nonumber\\
&&\hspace*{-5mm} \times\Big(\ln\frac{\Lambda}{\mu} + b_{2,13}\Big) +\frac{13}{24\pi^2}\mbox{tr}\Big(Y_U^+ Y_U\Big)\Big(\ln\frac{\Lambda}{\mu} + b_{2,1U}\Big)
+ \frac{7}{24\pi^2}\mbox{tr}\Big(Y_D^+ Y_D\Big)\Big(\ln\frac{\Lambda}{\mu} + b_{2,1D}\Big)
+ \frac{3}{8\pi^2}\nonumber\\
&&\hspace*{-5mm} \times \mbox{tr}\Big(Y_E^+ Y_E\Big)\Big(\ln\frac{\Lambda}{\mu} + b_{2,1E}\Big) \bigg] + O(\alpha^2,\alpha Y^2, Y^4);
\end{eqnarray}

\noindent
Similarly, the renormalization constants for the chiral matter superfields are given by the expressions

\begin{eqnarray}\label{Z_Q}
&&\hspace*{-5mm} (Z_Q)^T = 1 + \frac{\alpha_1}{60\pi}\Big(\ln\frac{\Lambda}{\mu} + g_{Q1}\Big) + \frac{3\alpha_2}{4\pi}\Big(\ln\frac{\Lambda}{\mu} + g_{Q2}\Big) + \frac{4\alpha_3}{3\pi}\Big(\ln\frac{\Lambda}{\mu} + g_{Q3}\Big) \nonumber\\
&& \hspace*{-5mm}\qquad\qquad\quad
- \frac{1}{8\pi^2} Y_{U} Y_{U}^+ \Big(\ln\frac{\Lambda}{\mu} + g_{QU}\Big) -\frac{1}{8\pi^2} Y_{D} Y_{D}^+ \Big(\ln\frac{\Lambda}{\mu} + g_{QD}\Big)+ O(\alpha^2,\alpha Y^2, Y^4);\\
&&\hspace*{-5mm} Z_U = 1 + \frac{4 \alpha_{1}}{15 \pi}\Big(\ln\frac{\Lambda}{\mu} + g_{U1}\Big) + \frac{4\alpha_{3}}{3\pi}\Big(\ln\frac{\Lambda}{\mu} + g_{U3}\Big)
\nonumber\\
&&\hspace*{-5mm} \qquad\qquad\qquad\qquad\qquad\qquad\qquad\qquad\ \
- \frac{1}{4\pi^2}\, Y_{U}^+ Y_{U}\Big(\ln\frac{\Lambda}{\mu} + g_{UU}\Big)+ O(\alpha^2,\alpha Y^2, Y^4);\\
&&\hspace*{-5mm} Z_D = 1 + \frac{\alpha_{1}}{15\pi}\Big(\ln\frac{\Lambda}{\mu} + g_{D1}\Big) + \frac{4\alpha_{3}}{3\pi}\Big(\ln\frac{\Lambda}{\mu} + g_{D3}\Big)
\nonumber\\
&&\hspace*{-5mm} \qquad\qquad\qquad\qquad\qquad\qquad\qquad\qquad\ \
- \frac{1}{4\pi^2}\, Y_{D}^+ Y_{D} \Big(\ln\frac{\Lambda}{\mu} + g_{DD}\Big)+ O(\alpha^2,\alpha Y^2, Y^4);\\
&&\hspace*{-5mm} (Z_L)^T = 1 + \frac{3\alpha_{1}}{20\pi}\Big(\ln\frac{\Lambda}{\mu} + g_{L1}\Big) + \frac{3\alpha_{2}}{4\pi}\Big(\ln\frac{\Lambda}{\mu} + g_{L2}\Big) \nonumber\\
&&\hspace*{-5mm} \qquad\qquad\qquad\qquad\qquad\qquad\qquad\qquad\ \
- \frac{1}{8\pi^2}\, Y_{E} Y_{E}^+ \Big(\ln\frac{\Lambda}{\mu} + g_{LE}\Big)+ O(\alpha^2,\alpha Y^2, Y^4);\\
&&\hspace*{-5mm} Z_E = 1 + \frac{3\alpha_{1}}{5\pi} \Big(\ln\frac{\Lambda}{\mu} + g_{E1}\Big) - \frac{1}{4\pi^2}\, Y_{E}^+ Y_{E} \Big(\ln\frac{\Lambda}{\mu} + g_{EE}\Big)+ O(\alpha^2,\alpha Y^2, Y^4);\\
&&\hspace*{-5mm} Z_{H_u} = 1 + \frac{3\alpha_{1}}{20\pi}\Big(\ln\frac{\Lambda}{\mu} + g_{H_u1}\Big) + \frac{3\alpha_{2}}{4\pi}\Big(\ln\frac{\Lambda}{\mu} + g_{H_u2}\Big)
\nonumber\\
&&\hspace*{-5mm} \qquad\qquad\qquad\qquad\qquad\qquad\qquad\
- \frac{3}{8\pi^2}\, \mbox{tr}\Big(Y_{U}^+ Y_{U}\Big)\Big(\ln\frac{\Lambda}{\mu} + g_{H_uU}\Big)+ O(\alpha^2,\alpha Y^2, Y^4);\qquad\\
\label{Z_Hd}
&&\hspace*{-5mm} Z_{H_d} = 1 + \frac{3\alpha_{1}}{20\pi}\Big(\ln\frac{\Lambda}{\mu} + g_{H_d1}\Big) + \frac{3\alpha_{2}}{4\pi}\Big(\ln\frac{\Lambda}{\mu} + g_{H_d2}\Big)\nonumber\\
&&\hspace*{-5mm} - \frac{3}{8\pi^2}\, \mbox{tr}\Big( Y_{D}^+ Y_{D}\Big) \Big(\ln\frac{\Lambda}{\mu} + g_{H_dD}\Big) - \frac{1}{8\pi^2} \mbox{tr}\Big(Y_{E}^+ Y_{E}\Big) \Big(\ln\frac{\Lambda}{\mu} + g_{H_dE}\Big)+ O(\alpha^2,\alpha Y^2, Y^4).
\end{eqnarray}

\noindent
Note that the coefficients of powers of $\ln\Lambda/\mu$ in the (one- and two-loop) terms written explicitly in Eqs. (\ref{Alpha3}) --- (\ref{Alpha1}) and in the (one-loop) terms written explicitly in (\ref{Z_Q}) --- (\ref{Z_Hd}) are scheme independent \cite{Martin:1993zk}, see also \cite{Korneev:2021zdz} for the detailed discussion. Certainly, it is important that the considered finite renormalizations relating various subtraction schemes are compatible with the structure of quantum corrections \cite{Jack:2016tpp}. (Otherwise, the two-loop contribution to the gauge $\beta$-functions would depend on the renormalization scheme \cite{McKeon:2017mjq} for theories with multiple gauge couplings.)

As a rule, due to the nonrenormalization of superpotential \cite{Grisaru:1979wc}, the renormalization of the Yukawa couplings in ${\cal N}=1$ supersymmetric theories is chosen according to the prescription

\begin{eqnarray}\label{Usual_Yukawa_Renormalization_U}
&& Y_{0U} = (Z_{H_u})^{-1/2} \left((Z_Q)^T\right)^{-1/2} Y_{U} (Z_U)^{-1/2};\\
\label{Usual_Yukawa_Renormalization_D}
&& Y_{0D} = (Z_{H_d})^{-1/2} \left((Z_Q)^T\right)^{-1/2} Y_{D} (Z_D)^{-1/2};\\
\label{Usual_Yukawa_Renormalization_E}
&& Y_{0E} = (Z_{H_d})^{-1/2} \left((Z_L)^T\right)^{-1/2} Y_{E} (Z_E)^{-1/2}.
\end{eqnarray}

\noindent
However, this prescription is not unique, because it is also possible to make finite renormalizations of the Yukawa couplings. Such renormalizations are very important, e.g., for constructing finite ${\cal N}=1$ supersymmetric theories, see \cite{Kazakov:1986bs,Ermushev:1986cu,Jack:1996qq}. That is why here we will use more general equations than Eq. (\ref{Usual_Yukawa_Renormalization_U}) --- (\ref{Usual_Yukawa_Renormalization_E}), namely,

\begin{eqnarray}
&&\hspace*{-5mm} Y_{0U} = \bigg[1 -\frac{13\alpha_1}{60\pi}\Big(\ln\frac{\Lambda}{\mu} + j_{U1}\Big) - \frac{3\alpha_2}{4\pi}\Big(\ln\frac{\Lambda}{\mu} + j_{U2}\Big) -\frac{4\alpha_3}{3\pi}\Big(\ln\frac{\Lambda}{\mu} + j_{U3}\Big) +\frac{3}{16\pi^2}\mbox{tr}\Big(Y_{U}^+ Y_{U}\Big) \nonumber\\
&&\hspace*{-5mm} \times \Big(\ln\frac{\Lambda}{\mu} + j_{UtU}\Big) +\frac{1}{16\pi^2} Y_{D} Y_{D}^+\Big(\ln\frac{\Lambda}{\mu} + j_{UD}\Big) +\frac{3}{16\pi^2} Y_{U} Y_{U}^+ \Big(\ln\frac{\Lambda}{\mu} + j_{UU}\Big) \bigg] Y_U \nonumber\\
&&\hspace*{-5mm} + O\Big(\alpha^2 Y,\alpha Y^3, Y^5\Big);\vphantom{\frac{1}{2}}\\
&&\hspace*{-5mm} Y_{0D} = \bigg[1 -\frac{7\alpha_1}{60\pi}\Big(\ln\frac{\Lambda}{\mu} + j_{D1}\Big) - \frac{3\alpha_2}{4\pi}\Big(\ln\frac{\Lambda}{\mu} + j_{D2}\Big) -\frac{4\alpha_3}{3\pi}\Big(\ln\frac{\Lambda}{\mu} + j_{D3}\Big) +\frac{3}{16\pi^2}\mbox{tr}\Big(Y_{D}^+ Y_{D}\Big) \nonumber\\
&&\hspace*{-5mm} \times \Big(\ln\frac{\Lambda}{\mu} + j_{DtD}\Big) +\frac{1}{16\pi^2}\mbox{tr}\Big(Y_{E}^+ Y_{E}\Big)
\Big(\ln\frac{\Lambda}{\mu} + j_{DtE}\Big) +\frac{3}{16\pi^2} Y_{D} Y_{D}^+\Big(\ln\frac{\Lambda}{\mu} + j_{DD}\Big) +\frac{1}{16\pi^2} Y_{U}\nonumber\\
&&\hspace*{-5mm} \times Y_{U}^+ \Big(\ln\frac{\Lambda}{\mu} + j_{DU}\Big) \bigg] Y_D + O\Big(\alpha^2 Y,\alpha Y^3, Y^5\Big);\\
&&\hspace*{-5mm} Y_{0E} = \bigg[1 -\frac{9\alpha_1}{20\pi}\Big(\ln\frac{\Lambda}{\mu} + j_{E1}\Big) - \frac{3\alpha_2}{4\pi}\Big(\ln\frac{\Lambda}{\mu} + j_{E2}\Big) +\frac{3}{16\pi^2}\mbox{tr}\Big(Y_{D}^+ Y_{D}\Big)  \Big(\ln\frac{\Lambda}{\mu} + j_{EtD}\Big) \nonumber\\
&&\hspace*{-5mm} +\frac{1}{16\pi^2} \mbox{tr}\Big(Y_{E}^+ Y_{E}\Big) \Big(\ln\frac{\Lambda}{\mu} + j_{EtE}\Big) +\frac{3}{16\pi^2} Y_{E} Y_{E}^+ \Big(\ln\frac{\Lambda}{\mu} + j_{EE}\Big) \bigg] Y_E + O\Big(\alpha^2 Y,\alpha Y^3, Y^5\Big).
\end{eqnarray}

\noindent
If the finite constants $j$ satisfy the equations

\begin{eqnarray}\label{J_Vs_G1}
&& j_{U1} = \frac{9}{26} g_{H_u1} + \frac{1}{26} g_{Q1} + \frac{8}{13} g_{U1};\qquad j_{U2} = \frac{1}{2} g_{H_u2} + \frac{1}{2} g_{Q2};\qquad j_{U3} = \frac{1}{2} g_{Q3} + \frac{1}{2} g_{U3};\nonumber\\
&& j_{UtU} = g_{H_uU};\qquad\ \, j_{UD} = g_{QD};\qquad\ \ j_{UU} = \frac{1}{3} g_{QU} + \frac{2}{3} g_{UU};\\
\label{J_Vs_G2}
&& j_{D1} = \frac{9}{14} g_{H_d1} + \frac{1}{14} g_{Q1} + \frac{2}{7} g_{D1};\qquad\ \, j_{D2} = \frac{1}{2} g_{H_d2} + \frac{1}{2} g_{Q2}; \quad\ \ \, j_{D3} = \frac{1}{2} g_{Q3} + \frac{1}{2} g_{D3};\nonumber\\
&& j_{DtD} = g_{H_dD};\qquad j_{DtE} = g_{H_dE};\qquad\ j_{DD} = \frac{1}{3} g_{QD} + \frac{2}{3} g_{DD};\quad\ \ j_{DU} = g_{QU};\\
\label{J_Vs_G3}
&& j_{E1} = \frac{1}{6} g_{H_d1} + \frac{1}{6} g_{L1} + \frac{2}{3} g_{E1};\qquad\quad\ \ j_{E2} = \frac{1}{2} g_{H_d2} + \frac{1}{2} g_{L2};\qquad\ j_{EtD} = g_{H_dD};\qquad \nonumber\\
&& j_{EtE} = g_{H_dE};\qquad\ j_{EE} = \frac{1}{3} g_{H_dE} + \frac{2}{3} g_{EE},
\end{eqnarray}

\noindent
then we obtain the prescription (\ref{Usual_Yukawa_Renormalization_U}) --- (\ref{Usual_Yukawa_Renormalization_E}). However, below we will not in general assume that these relations are satisfied.

RGFs defined in terms of the renormalized couplings depend on the finite constants specifying a renormalization prescription. The first scheme-dependent terms in the $\beta$-functions and in the anomalous dimensions of the matter superfields come from the three- and two-loop approximations, respectively. In this section we present expressions for the two-loop anomalous dimensions for all MSSM matter superfields defined in terms of the renormalized couplings. They were constructed from the anomalous dimensions defined in terms of the bare coupling by integrating the second equation in (\ref{RGFs_Bare_Definition}) and substituting the results into the second equation in (\ref{RGFs_Standard_Definition}). The result is written as

\begin{eqnarray}\label{MSSM_Anomalous_Dimentions_Q_Renormalized}
&&\hspace*{-5mm} \widetilde\gamma_Q(\alpha,Y)^T = - \frac{\alpha_1}{60\pi} - \frac{3\alpha_2}{4\pi} - \frac{4\alpha_3}{3\pi} + \frac{1}{8\pi^2}\Big(Y_{U} Y_{U}^+ + Y_{D} Y_{D}^+\Big)
+\frac{1}{2\pi^2}\bigg[\frac{\alpha_{1}^2}{3600} + \frac{9\alpha_{2}^2}{16}  + \frac{16\alpha_{3}^2}{9} \nonumber\\
&&\hspace*{-5mm} +\frac{\alpha_{1}\alpha_{2}}{40} + \frac{2\alpha_{1}\alpha_{3}}{45} + 2\alpha_{2}\alpha_{3}
+ \frac{11\alpha_{1}^2}{100} \Big(\ln a_{1}+1+\frac{A}{2}+g_{Q1}-b_{1,1}\Big)
+ \frac{3\alpha_{2}^2}{4}\Big(-6\ln a_{\varphi,2}\nonumber\\
&&\hspace*{-5mm} +7\ln a_{2}+1+\frac{A}{2} + g_{Q2}-b_{1,2}\Big)
- 4\alpha_{3}^2\Big(3\ln a_{\varphi,3}- 2\ln a_{3}+1+\frac{A}{2}+g_{Q3}-b_{1,3}\Big)\bigg]
+\frac{1}{8\pi^3}\nonumber\\
&&\hspace*{-5mm}\times Y_{U} Y_{U}^+ \bigg[\frac{\alpha_{1}}{5} + \frac{13\alpha_{1}}{60}\Big(B-A + 2g_{QU} - 2j_{U1}\Big) +\frac{3\alpha_{2}}{4}\Big(B-A+2g_{QU}-2j_{U2}\Big)
+ \frac{4\alpha_{3}}{3}\Big(B\nonumber\\
&&\hspace*{-5mm} -A+2g_{QU}-2j_{U3}\Big)\bigg]
+ \frac{1}{8\pi^3} Y_{D} Y_{D}^+ \bigg[\frac{\alpha_{1}}{10} + \frac{7\alpha_{1}}{60}\Big(B-A+2g_{QD}-2j_{D1}\Big) +\frac{3\alpha_{2}}{4}\Big(B-A\nonumber\\
&&\hspace*{-5mm} +2g_{QD}-2j_{D2}\Big) + \frac{4\alpha_{3}}{3} \Big(B-A+2g_{QD}-2j_{D3}\Big) \bigg] -\frac{1}{(8\pi^2)^2}\bigg[(Y_{U} Y_{U}^+)^2\Big(1+3g_{QU}-3j_{UU}\Big) \nonumber\\
&&\hspace*{-5mm} + (Y_{D} Y_{D}^+)^2\Big(1+3g_{QD}-3j_{DD}\Big) + \frac{1}{2}\Big(Y_D Y_D^+ Y_U Y_U^+ + Y_U Y_U^+ Y_D Y_D^+\Big)\Big(g_{QD}+g_{QU}-j_{UD}\nonumber\\
&&\hspace*{-5mm} -j_{DU}\Big) + \frac{3}{2} Y_{U} Y_{U}^+\, \mbox{tr}\Big(Y_{U} Y_{U}^+\Big)\Big(1+2g_{QU}-2j_{UtU}\Big) + \frac{3}{2} Y_{D} Y_{D}^+\, \mbox{tr}\Big(Y_{D} Y_{D}^+\Big)\Big(1+2g_{QD}\nonumber\\
&&\hspace*{-5mm} -2j_{DtD}\Big) + \frac{1}{2} Y_{D} Y_{D}^+\, \mbox{tr}\Big(Y_{E} Y_{E}^+\Big)\Big(1+2g_{QD}-2j_{DtE}\Big) \bigg] + O\Big(\alpha^3,\alpha^2 Y^2, \alpha Y^4, Y^6\Big);\qquad\\
&&\hspace*{-5mm} \widetilde\gamma_U(\alpha,Y) = - \frac{4 \alpha_{1}}{15 \pi} - \frac{4\alpha_{3}}{3\pi} + \frac{1}{4\pi^2}\, Y_{U}^+ Y_{U}
+ \frac{1}{2\pi^2}\bigg[\frac{16\alpha_{1}^2}{225} + \frac{16\alpha_{3}^2}{9} + \frac{32\alpha_{1}\alpha_{3}}{45} + \frac{44\alpha_{1}^2}{25}\Big(\ln a_{1} \nonumber\\
&&\hspace*{-5mm} +1+\frac{A}{2}+g_{U1}-b_{1,1}\Big) - 4\alpha_{3}^2\Big(3\ln a_{\varphi,3} - 2\ln a_{3}+1+\frac{A}{2}+g_{U3} - b_{1,3}\Big)\bigg] + \frac{1}{8\pi^3} Y_{U}^+ Y_{U} \nonumber\\
&&\hspace*{-5mm} \times\bigg[-\frac{\alpha_{1}}{10} + \frac{3\alpha_{2}}{2} + \frac{13\alpha_{1}}{30}\Big(B-A+2g_{UU}-2j_{U1}\Big) + \frac{3\alpha_{2}}{2}\Big(B-A+2g_{UU}-2j_{U2}\Big)
+ \frac{8\alpha_{3}}{3}\nonumber\\
&&\hspace*{-5mm} \times \Big(B-A+2g_{UU}-2j_{U3}\Big)\bigg] - \frac{1}{(8\pi^2)^2}\bigg[(Y_{U}^+ Y_{U})^2\Big(1+6g_{UU}-6j_{UU}\Big) + 3 Y_{U}^+Y_{U}\, \mbox{tr}\Big(Y_{U} Y_{U}^+\Big)\nonumber\\
&&\hspace*{-5mm} \times \Big(1+2g_{UU}-2j_{UtU}\Big) + Y_{U}^+ Y_{D} Y_{D}^+ Y_{U}\Big(1+2g_{UU}-2j_{UD}\Big) \bigg] + O\Big(\alpha^3,\alpha^2 Y^2,\alpha Y^4, Y^6\Big);\\
&&\hspace*{-5mm} \widetilde\gamma_D(\alpha,Y) = - \frac{\alpha_{1}}{15\pi} - \frac{4\alpha_{3}}{3\pi} + \frac{1}{4\pi^2}\, Y_{D}^+ Y_{D}
+ \frac{1}{2\pi^2}\bigg[\frac{\alpha_{1}^2}{225} + \frac{16\alpha_{3}^2}{9} + \frac{8\alpha_{1}\alpha_{3}}{45} + \frac{11\alpha_{1}^2}{25}\Big(\ln a_{1}+1\nonumber\\
&&\hspace*{-5mm} +\frac{A}{2}+g_{D1}-b_{1,1}\Big) -4\alpha_{3}^2\Big(3\ln a_{\varphi,3}-2\ln a_{3}+1+\frac{A}{2}+ g_{D3}-b_{1,3}\Big)\bigg] + \frac{1}{8\pi^3} Y_{D}^+ Y_{D}\bigg[\frac{\alpha_{1}}{10}
\nonumber\\
&&\hspace*{-5mm}  + \frac{3\alpha_{2}}{2} + \frac{7\alpha_{1}}{30}\Big(B-A+2g_{DD}-2j_{D1}\Big) + \frac{3\alpha_{2}}{2}\Big(B-A+2g_{DD}-2j_{D2}\Big) + \frac{8\alpha_{3}}{3}\Big(B-A\nonumber\\
&&\hspace*{-5mm} +2g_{DD}-2j_{D3}\Big)\bigg] - \frac{1}{(8\pi^2)^2}\bigg[(Y_{D}^+ Y_{D})^2
\Big(1 +6g_{DD}-6j_{DD}\Big) + 3 Y_{D}^+Y_{D}\, \mbox{tr}\Big(Y_{D} Y_{D}^+\Big) \Big(1+2g_{DD} \nonumber\\
&&\hspace*{-5mm} -2j_{DtD}\Big) +  Y_{D}^+Y_{D}\, \mbox{tr} \Big(Y_{E} Y_{E}^+\Big)\Big(1+2g_{DD}-2j_{DtE}\Big) + Y_{D}^+ Y_{U} Y_{U}^+ Y_{D}\Big(1 + 2g_{DD} - 2j_{DU}\Big) \bigg]\nonumber\\
&&\hspace*{-5mm} + O\Big(\alpha^3,\alpha^2 Y^2,\alpha Y^4, Y^6\Big);\vphantom{\frac{1}{2}}\\
&&\hspace*{-5mm} \widetilde\gamma_L(\alpha,Y)^T = - \frac{3\alpha_{1}}{20\pi} - \frac{3\alpha_{2}}{4\pi} + \frac{1}{8\pi^2}\, Y_{E} Y_{E}^+
+ \frac{1}{2\pi^2}\bigg[ \frac{9\alpha_{1}^2}{400} + \frac{9\alpha_{2}^2}{16} + \frac{9\alpha_{1}\alpha_{2}}{40} + \frac{99\alpha_{1}^2}{100} \Big(\ln a_{1}  \nonumber\\
&&\hspace*{-5mm} + 1+\frac{A}{2} + g_{L1}-b_{1,1}\Big)
+ \frac{3\alpha_{2}^2}{4} \Big(-6\ln a_{\varphi,2} + 7\ln a_{2} + 1 +\frac{A}{2} + g_{L2} - b_{1,2}\Big)\bigg] + \frac{1}{8\pi^3} Y_{E} Y_{E}^+  \nonumber\\
&&\hspace*{-5mm} \times \bigg[\frac{3\alpha_{1}}{10} + \frac{9\alpha_{1}}{20}\Big(B-A +2g_{LE}-2j_{E1}\Big) + \frac{3\alpha_{2}}{4}\Big(B-A + 2g_{LE} - 2j_{E2}\Big)\bigg]
- \frac{1}{(8\pi^2)^2} \nonumber\\
&&\hspace*{-5mm} \times \bigg[(Y_{E} Y_{E}^+)^2\Big(1+3g_{LE} - 3j_{EE}\Big) + \frac{3}{2} Y_{E} Y_{E}^+\, \mbox{tr} \Big( Y_{D}^+ Y_{D}\Big)\Big(1 + 2g_{LE} - 2 j_{EtD}\Big) + \frac{1}{2} Y_{E} Y_{E}^+
\nonumber\\
&& \hspace*{-5mm} \times\, \mbox{tr} \Big(Y_{E}^+ Y_{E}\Big)\Big(1+2g_{LE}-2j_{EtE}\Big)\bigg] + O\Big(\alpha^3,\alpha^2 Y^2,\alpha Y^4, Y^6\Big);\\
&&\hspace*{-5mm} \widetilde\gamma_E(\alpha,Y) = - \frac{3\alpha_{1}}{5\pi} + \frac{1}{4\pi^2}\, Y_{E}^+ Y_{E} + \frac{1}{2\pi^2}\cdot \frac{9\alpha_{1}^2}{25} \bigg[1+11 \Big(\ln a_1+1+\frac{A}{2}+g_{E1}-b_{1,1}\Big)\bigg]
\nonumber\\
&&\hspace*{-5mm} + \frac{1}{8\pi^3} Y_{E}^+ Y_{E} \bigg[-\frac{3\alpha_{1}}{10} + \frac{3\alpha_{2}}{2} + \frac{9\alpha_{1}}{10} \Big(B-A+2g_{EE}-2j_{E1}\Big) + \frac{3\alpha_{2}}{2} \Big(B-A+2g_{EE}
\nonumber\\
&&\hspace*{-5mm} -2j_{E2}\Big)\bigg] - \frac{1}{(8\pi^2)^2}\bigg[(Y_{E}^+ Y_{E})^2\Big(1+6g_{EE}-6j_{EE}\Big) + 3 Y_{E}^+ Y_{E}\, \mbox{tr} \Big( Y_{D}^+ Y_{D}\Big)\Big(1 +2 g_{EE} - 2j_{EtD}\Big) \nonumber\\
&&\hspace*{-5mm} +  Y_{E}^+ Y_{E}\,\mbox{tr} \Big(Y_{E}^+ Y_{E}\Big)\Big(1 + 2g_{EE}-2j_{EtE}\Big)\bigg] + O\Big(\alpha^3,\alpha^2 Y^2,\alpha Y^4, Y^6\Big);\\
&&\hspace*{-5mm} \widetilde\gamma_{H_u}(\alpha,Y) = - \frac{3\alpha_{1}}{20\pi} - \frac{3\alpha_{2}}{4\pi} + \frac{3}{8\pi^2}\, \mbox{tr}\Big(Y_{U}^+ Y_{U}\Big) + \frac{1}{2\pi^2}\bigg[ \frac{9\alpha_{1}^2}{400} + \frac{9\alpha_{2}^2}{16} + \frac{9\alpha_{1}\alpha_{2}}{40} + \frac{99\alpha_{1}^2}{100} \Big(\ln a_{1}\nonumber\\
&&\hspace*{-5mm} + 1+\frac{A}{2} +g_{H_u1} - b_{1,1}\Big) + \frac{3\alpha_{2}^2}{4} \Big(-6\ln a_{\varphi,2}+ 7\ln a_{2} + 1+\frac{A}{2} +g_{H_u2} - b_{1,2}\Big)\bigg]
\nonumber\\
&&\hspace*{-5mm} +\frac{1}{8\pi^3} \mbox{tr}\Big(Y_{U} Y_{U}^+\Big) \bigg[\frac{\alpha_{1}}{5} + 4\alpha_{3} + \frac{13\alpha_{1}}{20}\Big(B-A+2g_{H_uU}-2j_{U1}\Big) + \frac{9\alpha_{2}}{4}\Big(B-A+2g_{H_uU}
\nonumber\\
&&\hspace*{-5mm} -2j_{U2}\Big) + 4\alpha_{3}\Big(B-A+2g_{H_uU}-2j_{U3}\Big)\bigg]
-\frac{1}{(8\pi^2)^2}\bigg[\frac{3}{2}\,\mbox{tr}\Big(Y_{D} Y_{D}^+ Y_{U} Y_{U}^+\Big)\Big(1+2g_{H_uU} \nonumber\\
&&\hspace*{-5mm} -2j_{UD}\Big) + \frac{9}{2}\,\mbox{tr}\Big((Y_{U}Y_{U}^+)^2\Big)\Big(1+2g_{H_uU}-2j_{UU}\Big) + 9 \Big[\mbox{tr} \Big(Y_U^+ Y_U\Big)\Big]^2\Big(g_{H_uU}-j_{UtU}\Big)\bigg]
\nonumber\\
&&\hspace*{-5mm} + O\Big(\alpha^3,\alpha^2 Y^2,\alpha Y^4, Y^6\Big);\\
\label{MSSM_Anomalous_Dimentions_Hd_Renormalized}
&&\hspace*{-5mm} \widetilde\gamma_{H_d}(\alpha,Y) = - \frac{3\alpha_{1}}{20\pi} - \frac{3\alpha_{2}}{4\pi} + \frac{1}{8\pi^2}\, \mbox{tr}\Big(3\, Y_{D}^+ Y_{D} + Y_{E}^+ Y_{E}\Big)  + \frac{1}{2\pi^2}\bigg[ \frac{9\alpha_{1}^2}{400} + \frac{9\alpha_{2}^2}{16} + \frac{9}{40} \alpha_{1}\alpha_{2} \nonumber\\
&&\hspace*{-5mm}  + \frac{99\alpha_{1}^2}{100} \Big(\ln a_{1} + 1+\frac{A}{2} +g_{H_d1}-b_{1,1}\Big)
+ \frac{3\alpha_{2}^2}{4} \Big(-6\ln a_{\varphi,2} +7\ln a_{2} + 1 +\frac{A}{2} +g_{H_d2} - b_{1,2}\Big)\bigg]\nonumber\\
&&\hspace*{-5mm}  + \frac{1}{8\pi^3}\mbox{tr}\Big(Y_{E} Y_{E}^+\Big)\bigg[\frac{3\alpha_{1}}{10} + \frac{9\alpha_{1}}{20}\Big(B-A+2g_{H_dE}-2j_{E1}\Big) + \frac{3\alpha_{2}}{4}\Big(B-A+2g_{H_dE}-2j_{E2}\Big)\bigg]
\nonumber\\
&&\hspace*{-5mm} + \frac{1}{8\pi^3} \mbox{tr}\Big(Y_{D} Y_{D}^+\Big)\bigg[-\frac{\alpha_{1}}{10} + 4\alpha_{3} + \frac{7\alpha_{1}}{20}\Big(B-A+2g_{H_dD}-2j_{D1}\Big) + \frac{9\alpha_{2}}{4}\Big(B-A+2g_{H_dD}
\nonumber\\
&&\hspace*{-5mm} -2j_{D2}\Big) + 4\alpha_{3}\Big(B-A+2g_{H_dD}-2j_{D3}\Big)\bigg] - \frac{1}{(8\pi^2)^2}\bigg[\frac{3}{2}\mbox{tr}\Big((Y_{E} Y_{E}^+)^2\Big)\Big(1+2g_{H_dE}-2j_{EE}\Big)\nonumber\\
&&\hspace*{-5mm} + \frac{3}{2}\mbox{tr}\Big(Y_{D} Y_{D}^+ Y_{U} Y_{U}^+\Big)\Big(1+2g_{H_dD}-2j_{DU}\Big) + \frac{9}{2} \mbox{tr}\Big((Y_{D} Y_{D}^+)^2\Big)\Big(1+2g_{H_dD}-2j_{DD}\Big) \nonumber\\
&&\hspace*{-5mm} + 9\Big[\mbox{tr}\Big(Y_D^+ Y_D\Big)\Big]^2 \Big(g_{H_dD}-j_{DtD}\Big) + 3\, \mbox{tr}\Big(Y_D^+ Y_D\Big)\, \mbox{tr}\Big(Y_E^+ Y_E\Big) \Big(g_{H_dD} + g_{H_dE} -j_{DtE} - j_{EtD}\Big)\nonumber\\
&&\hspace*{-5mm} +  \Big[\mbox{tr}\Big(Y_E^+ Y_E\Big)\Big]^2 \Big(g_{H_dE}-j_{EtE}\Big) \bigg]
+ O\Big(\alpha^3,\alpha^2 Y^2,\alpha Y^4, Y^6\Big).
\end{eqnarray}

In the HD+MSL scheme all finite constants $b$, $g$, and $j$ vanish by construction \cite{Kataev:2013eta,Shakhmanov:2017wji,Stepanyantz:2017sqg}, and these RGFs coincide with the ones defined in terms of the bare couplings (given by Eqs. (\ref{MSSM_Anomalous_Dimentions_Q_Bare}) --- (\ref{MSSM_Anomalous_Dimentions_Hd_Bare})) after the formal replacement of the arguments $\alpha \to \alpha_0$, $Y\to Y_0$.

\section{Three-loop MSSM $\beta$-functions from the NSVZ equations}
\hspace*{\parindent}\label{Section_MSSM_Beta_Functions}

In the case of using the higher covariant derivative regularization the three-loop $\beta$-function defined in terms of the bare couplings for MSSM can be found with the help of the NSVZ equations (\ref{MSSM_Exact_Beta3}) --- (\ref{MSSM_Exact_Beta1}), in which we should substitute the the two-loop expressions (\ref{MSSM_Anomalous_Dimentions_Q_Bare}) --- (\ref{MSSM_Anomalous_Dimentions_Hd_Bare}) for the anomalous dimensions of all chiral matter superfields. The result is presented in Appendix \ref{Appendix_Three_Loop_Beta}. Next, it is necessary to find RGFs defined in terms of the renormalized couplings. For this purpose we integrate the renormalization group equations (\ref{RGFs_Bare_Definition}) and find the dependence of all $\alpha_{0}$ on $\alpha$ and $Y$. Solving the resulting equations for the renormalized gauge couplings and substituting them into Eq. (\ref{RGFs_Standard_Definition}) we obtain the standard gauge $\beta$-functions

\begin{eqnarray}\label{Beta3_Renormalized}
&&\hspace*{-5mm} \frac{\widetilde\beta_3(\alpha,Y)}{\alpha_{3}^2} = - \frac{1}{2\pi} \bigg\{3 -\frac{11\alpha_{1}}{20\pi} -\frac{9\alpha_{2}}{4\pi} -\frac{7\alpha_{3}}{2\pi}
+ \frac{1}{8\pi^2}\, \mbox{tr}\Big(2\, Y_{U}^+ Y_{U} + 2\, Y_{D}^+ Y_{D}\Big) + \frac{1}{2\pi^2}\bigg[ \frac{137\alpha_{1}^2}{1200}\nonumber\\
&&\hspace*{-5mm}  + \frac{27\alpha_{2}^2}{16} + \frac{\alpha_{3}^2}{6} + \frac{3\alpha_{1}\alpha_{2}}{40} - \frac{11\alpha_{1}\alpha_{3}}{60}
-\frac{3\alpha_{2}\alpha_{3}}{4} + \frac{363\alpha_{1}^2}{100}\Big(\ln a_{1} + 1 +\frac{A}{2} + b_{2,31} - b_{1,1}\Big) + \frac{9\alpha_{2}^2}{4}\nonumber\\
&&\hspace*{-5mm} \times \Big(-6\ln a_{\varphi,2} + 7\ln a_{2} + 1 +\frac{A}{2} + b_{2,32} - b_{1,2} \Big)
- 24\alpha_{3}^2 \Big(3\ln a_{\varphi,3} -2\ln a_3 + 1 +\frac{A}{2} + \frac{7}{16} b_{2,33} \nonumber\\
&&\hspace*{-5mm} - \frac{7}{16} b_{1,3} \Big) \bigg] + \frac{1}{8\pi^3}\mbox{tr}\Big(Y_{U} Y_{U}^+\Big)\bigg[\frac{3\alpha_{1}}{20} + \frac{3\alpha_{2}}{4} + 3\alpha_{3} + \frac{13\alpha_{1}}{30} \Big(B-A+2b_{2,3U} -2j_{U1}\Big)
+ \frac{3\alpha_{2}}{2}
\nonumber\\
&&\hspace*{-5mm} \times \Big(B-A+2b_{2,3U}-2j_{U2}\Big) + \frac{8\alpha_{3}}{3}\Big(B-A+2b_{2,3U}-2j_{U3}\Big)\bigg] + \frac{1}{8\pi^3}\mbox{tr}\Big(Y_{D} Y_{D}^+\Big)\bigg[\frac{3\alpha_{1}}{20} + \frac{3\alpha_{2}}{4}
\nonumber\\
&&\hspace*{-5mm} + 3\alpha_{3} + \frac{7\alpha_{1}}{30} \Big(B-A+2b_{2,3D}-2j_{D1}\Big) + \frac{3\alpha_{2}}{2} \Big(B-A+2b_{2,3D}-2j_{D2}\Big) + \frac{8\alpha_{3}}{3}\Big(B-A\nonumber\\
&&\hspace*{-5mm} +2b_{2,3D}-2j_{D3}\Big)\bigg] - \frac{1}{(8\pi^2)^2}\bigg[\, \frac{3}{2}\mbox{tr}\Big((Y_{U} Y_{U}^+)^2\Big)\Big(1+4b_{2,3U}-4j_{UU}\Big) + \frac{3}{2}\mbox{tr}\Big((Y_{D} Y_{D}^+)^2\Big)\Big(1\nonumber\\
&&\hspace*{-5mm} +4b_{2,3D}-4j_{DD}\Big) + 3\Big(\mbox{tr}(Y_{U} Y_{U}^+)\Big)^2\Big(1+2b_{2,3U}-2j_{UtU}\Big)  + 3\Big(\mbox{tr}(Y_{D} Y_{D}^+)\Big)^2\Big(1+2b_{2,3D}\nonumber\\
&&\hspace*{-5mm} -2j_{DtD}\Big) + \mbox{tr}\Big(Y_{E} Y_{E}^+\Big)\,\mbox{tr}\Big(Y_{D} Y_{D}^+\Big)\Big(1+2b_{2,3D}-2j_{DtE}\Big) + \mbox{tr}\Big(Y_{D} Y_{D}^+ Y_{U} Y_{U}^+\Big)\Big(1+2b_{2,3U}
\nonumber\\
&&\hspace*{-5mm}
+ 2b_{2,3D} - 2j_{UD}-2j_{DU}\Big)\bigg]\bigg\}
+ O(\alpha^3,\alpha^2 Y^2, \alpha Y^4, Y^6);\qquad\\
&&\vphantom{1}\nonumber\\
\label{Beta2_Renormaluzed}
&&\hspace*{-5mm} \frac{\widetilde\beta_2(\alpha,Y)}{\alpha_{2}^2} = - \frac{1}{2\pi} \bigg\{-1  -\frac{9\alpha_{1}}{20\pi}- \frac{25\alpha_{2}}{4\pi} -\frac{6\alpha_{3}}{\pi} + \frac{1}{8\pi^2}\,
\mbox{tr}\Big(3\, Y_{U}^+ Y_{U} + 3\, Y_{D}^+ Y_{D} + Y_{E}^+ Y_{E} \Big) + \frac{1}{2\pi^2}\nonumber\\
&&\hspace*{-5mm} \times\bigg[\frac{23\alpha_{1}^2}{400} - \frac{137\alpha_{2}^2}{16} + 8 \alpha_{3}^2 - \frac{9\alpha_{1}\alpha_{2}}{40} + \frac{\alpha_{1}\alpha_{3}}{5} - 3\alpha_{2}\alpha_{3}
+ \frac{297\alpha_{1}^2}{100}\Big(\ln a_{1} + 1 + \frac{A}{2} + b_{2,21} - b_{1,1}\Big) \nonumber\\
&&\hspace*{-5mm}  + \frac{21\alpha_{2}^2}{4}\Big(7\ln a_{2} -6\ln a_{\varphi,2} + 1 + \frac{A}{2} + \frac{25}{21} b_{2,22} - \frac{25}{21} b_{1,2}\Big) -18\alpha_{3}^2 \Big(3\ln a_{\varphi,3} - 2\ln a_{3} + 1 + \frac{A}{2}\nonumber\\
&&\hspace*{-5mm}  + b_{2,23} - b_{1,3}\Big)\bigg]
+ \frac{1}{8\pi^3}\mbox{tr}\Big(Y_{U} Y_{U}^+\Big)\bigg[\,\frac{2\alpha_{1}}{5} + 3\alpha_{2} + 2\alpha_{3} + \frac{13\alpha_{1}}{20}\Big(B-A+2b_{2,2U}-2j_{U1}\Big) + \frac{9\alpha_{2}}{4}\nonumber\\
&&\hspace*{-5mm} \times\Big(B-A+2b_{2,2U}-2j_{U2}\Big) + 4\alpha_{3}\Big(B-A+2b_{2,2U}-2j_{U3}\Big)\bigg]
+ \frac{1}{8\pi^3}\mbox{tr}\Big(Y_{D} Y_{D}^+\Big)\bigg[\,\frac{\alpha_{1}}{10}+3\alpha_{2}\nonumber\\
&&\hspace*{-5mm} +2\alpha_{3} + \frac{7\alpha_{1}}{20}\Big(B-A+2b_{2,2D}-2j_{D1}\Big) + \frac{9\alpha_{2}}{4}\Big(B-A+2b_{2,2D}-2j_{D2}\Big) + 4\alpha_{3}\Big(B-A+2b_{2,2D}\nonumber\\
&&\hspace*{-5mm} -2j_{D3}\Big)\bigg] + \frac{1}{8\pi^3}\mbox{tr}\Big(Y_{E} Y_{E}^+\Big)\bigg[\,\frac{3\alpha_{1}}{10} + \alpha_{2} + \frac{9\alpha_{1}}{20}\Big(B-A+2b_{2,2E}-2j_{E1}\Big) + \frac{3\alpha_{2}}{4}\Big(B-A+2b_{2,2E}\nonumber\\
&&\hspace*{-5mm} -2j_{E2}\Big)\bigg]
- \frac{1}{(8\pi^2)^2}\bigg[\, \frac{15}{4}\,\mbox{tr}\Big((Y_{U} Y_{U}^+)^2\Big)\Big(1+\frac{12}{5} b_{2,2U} - \frac{12}{5} j_{UU}\Big) + \frac{15}{4}\,\mbox{tr}\Big((Y_{D} Y_{D}^+)^2\Big)\Big(1+\frac{12}{5} b_{2,2D}
\nonumber\\
&&\hspace*{-5mm} - \frac{12}{5} j_{DD}\Big)  + \frac{5}{4}\,\mbox{tr}\Big((Y_{E} Y_{E}^+)^2\Big) \Big(1+\frac{12}{5} b_{2,2E} - \frac{12}{5} j_{EE}\Big)
+ \frac{9}{4}\Big(\mbox{tr}(Y_{U} Y_{U}^+)\Big)^2 \Big(1+4b_{2,2U}-4j_{UtU}\Big)\nonumber\\
&&\hspace*{-5mm} + \frac{9}{4}\Big(\mbox{tr}(Y_{D} Y_{D}^+)\Big)^2  \Big(1+4b_{2,2D}-4j_{DtD}\Big)
+ \frac{1}{4}\Big(\mbox{tr}(Y_{E} Y_{E}^+)\Big)^2 \Big(1+4b_{2,2E}-4j_{EtE}\Big)
+ \frac{3}{2} \mbox{tr}\Big(Y_{E} Y_{E}^+\Big)
\nonumber\\
&&\hspace*{-5mm} \times\,\mbox{tr}\Big(Y_{D} Y_{D}^+\Big)\Big(1+2b_{2,2D}+2b_{2,2E}-2j_{EtD}-2j_{DtE}\Big)
+ \frac{3}{2}\mbox{tr}\Big(Y_{D} Y_{D}^+ Y_{U} Y_{U}^+\Big)\Big(1+2b_{2,2U}+2b_{2,2D}
\nonumber\\
&&\hspace*{-5mm} -2j_{UD}-2j_{DU}\Big) \bigg] \bigg\} + O(\alpha^3,\alpha^2 Y^2, \alpha Y^4, Y^6);\\
&&\vphantom{1}\nonumber\\
\label{Beta1_Renormalized}
&&\hspace*{-5mm} \frac{\widetilde\beta_1(\alpha,Y)}{\alpha_{1}^2} = - \frac{1}{2\pi}\cdot \frac{3}{5} \bigg\{ -11 -\frac{199\alpha_{1}}{60\pi} -\frac{9\alpha_{2}}{4\pi} -\frac{22\alpha_{3}}{3\pi}
+ \frac{1}{8\pi^2} \mbox{tr}\Big(\frac{13}{3} Y_{U}^+ Y_{U} + \frac{7}{3} Y_{D}^+ Y_{D} + 3 Y_{E}^+ Y_{E} \Big) \nonumber\\
&&\hspace*{-5mm}
+ \frac{1}{2\pi^2}\bigg[\, \frac{5131\alpha_{1}^2}{3600} + \frac{27\alpha_{2}^2}{16} + \frac{88\alpha_{3}^2}{9} + \frac{23\alpha_{1}\alpha_{2}}{40} + \frac{137\alpha_{1}\alpha_{3}}{45} + \alpha_{2}\alpha_{3}
+ \frac{2189\alpha_{1}^2}{100}\Big(\ln a_{1}+1+\frac{A}{2}+b_{2,11}
\nonumber\\
&&\hspace*{-5mm} - b_{1,1}\Big) + \frac{9\alpha_{2}^2}{4}\Big(7\ln a_{2}-6\ln a_{\varphi,2}+1+\frac{A}{2} + b_{2,12}-b_{1,2}\Big) -22 \alpha_{3}^2 \Big(3\ln a_{\varphi,3}-2\ln a_{3}+1+\frac{A}{2}
\nonumber\\
&&\hspace*{-5mm}
+b_{2,13}-b_{1,3}\Big) \bigg] + \frac{1}{8\pi^3}\mbox{tr}\Big(Y_{U} Y_{U}^+\Big)\bigg[\,2\alpha_{2} + 2\alpha_{3} + \frac{169\alpha_{1}}{180} \Big(B-A+2 b_{2,1U}-2j_{U1}\Big) + \frac{13\alpha_{2}}{4} \Big(B
\nonumber\\
&&\hspace*{-5mm}
-A+2 b_{2,1U}-2j_{U2}\Big) + \frac{52\alpha_{3}}{9} \Big(B-A+2 b_{2,1U}-2j_{U3}\Big)\bigg]
+ \frac{1}{8\pi^3}\mbox{tr}\Big(Y_{D} Y_{D}^+\Big)\bigg[\,\frac{\alpha_{2}}{2} + 2\alpha_{3} + \frac{49\alpha_{1}}{180} \nonumber\\
&&\hspace*{-5mm} \times\Big(B-A+2 b_{2,1D}-2j_{D1}\Big)
+ \frac{7\alpha_{2}}{4} \Big(B-A+2 b_{2,1D}-2j_{D2}\Big) + \frac{28\alpha_{3}}{9} \Big(B-A+2 b_{2,1D}-2j_{D3}\Big)\bigg]
\nonumber\\
&&\hspace*{-5mm}
+ \frac{1}{8\pi^3}\mbox{tr}\Big(Y_{E} Y_{E}^+\Big)\bigg[\,\frac{3\alpha_{2}}{2} + \frac{27\alpha_{1}}{20} \Big(B-A+2 b_{2,1E}-2j_{E1}\Big) + \frac{9\alpha_{2}}{4} \Big(B-A+2 b_{2,1E}-2j_{E2}\Big)\bigg]
\nonumber\\
&&\hspace*{-5mm}
- \frac{1}{(8\pi^2)^2}\bigg[\,\frac{15}{4} \mbox{tr}\Big((Y_{U} Y_{U}^+)^2\Big)\Big(1+\frac{52}{15} b_{2,1U} - \frac{52}{15} j_{UU}\Big) + \frac{11}{4} \mbox{tr}\Big((Y_{D} Y_{D}^+)^2\Big)\Big(1+\frac{28}{11}b_{2,1D} - \frac{28}{11} j_{DD}\Big)\nonumber\\
&&\hspace*{-5mm} + \frac{9}{4} \mbox{tr}\Big((Y_{E} Y_{E}^+)^2\Big)\Big(1 +4 b_{2,1E} - 4 j_{EE}\Big) + \frac{19}{6}\mbox{tr}\Big(Y_{D} Y_{D}^+ Y_{U} Y_{U}^+\Big)\Big(1 + \frac{26}{19} b_{2,1U} - \frac{26}{19} j_{UD} + \frac{14}{19} b_{2,1D} \nonumber\\
&&\hspace*{-5mm} - \frac{14}{19} j_{DU}\Big)  + \frac{17}{4} \Big(\mbox{tr}(Y_{U} Y_{U}^+)\Big)^2 \Big(1+\frac{52}{17} b_{2,1U} -\frac{52}{17}j_{UtU}\Big) + \frac{5}{4} \Big(\mbox{tr}(Y_{D} Y_{D}^+)\Big)^2 \Big(1+\frac{28}{5} b_{2,1D} - \frac{28}{5}\nonumber\\
&&\hspace*{-5mm} \times j_{DtD}\Big) + \frac{5}{4} \Big(\mbox{tr}(Y_{E} Y_{E}^+)\Big)^2 \Big(1+\frac{12}{5} b_{2,1E} - \frac{12}{5} j_{EtE}\Big)
+ \frac{25}{6} \mbox{tr}\Big(Y_{E} Y_{E}^+\Big)\,\mbox{tr}\Big(Y_{D} Y_{D}^+\Big)\Big(1+\frac{14}{25}b_{2,1D}\nonumber\\
&&\hspace*{-5mm} -\frac{14}{25}j_{DtE}+\frac{54}{25}b_{2,1E}-\frac{54}{25}j_{EtD}\Big)  \bigg]\bigg\} + O(\alpha^3,\alpha^2 Y^2, \alpha Y^4, Y^6).\vphantom{\frac{1}{2}}
\end{eqnarray}

The three-loop contributions to these $\beta$-functions depend on the finite constants which fix a renormalization prescription in the one- and two-loop approximations. As we already mentioned, the one- and two-loop contributions are scheme-independent in agreement with \cite{Martin:1993zk}.

In the HD+MSL scheme all finite constants are equal to 0, and RGFs defined in terms of the bare and renormalized couplings coincide up to a formal renaming of the arguments. In particular, this implies that the HD+MSL scheme is NSVZ.

\section{$\overline{\mbox{DR}}$ scheme}
\hspace*{\parindent}\label{Section_DR}

As a correctness test, we can verify that for certain values of the finite constants (satisfying Eqs. (\ref{J_Vs_G1}) --- (\ref{J_Vs_G3})) the expressions (\ref{MSSM_Anomalous_Dimentions_Q_Renormalized}) --- (\ref{MSSM_Anomalous_Dimentions_Hd_Renormalized}) for the two-loop anomalous dimensions and the expressions (\ref{Beta3_Renormalized}) --- (\ref{Beta1_Renormalized}) for the three-loop $\beta$-functions reproduce the known results in the $\overline{\mbox{DR}}$ scheme \cite{Jack:2004ch}. For completeness, in our notation they are presented in Appendix \ref{Appendix_RGFs_DR} and are given by Eqs. (\ref{MSSM_Anomalous_Dimentions_Q_DR}) --- (\ref{MSSM_Anomalous_Dimentions_Hd_DR}) and (\ref{MSSM_Beta3_DR}) --- (\ref{MSSM_Beta1_DR}). Note that this test is not trivial because NSVZ equations usually lead to some scheme independent consequences \cite{Kataev:2013csa,Kataev:2014gxa} which should be satisfied for all renormalization prescriptions. In particular, for MSSM from Eqs. (\ref{MSSM_Anomalous_Dimentions_Q_Renormalized}) --- (\ref{MSSM_Anomalous_Dimentions_Hd_Renormalized}) and Eqs. (\ref{Beta3_Renormalized}) --- (\ref{Beta1_Renormalized}) we see that the terms in the two-loop anomalous dimensions and in the three-loop $\beta$-functions proportional to $\alpha_1\alpha_2$, $\alpha_1\alpha_3$, and $\alpha_2\alpha_3$ are scheme independent and should be the same for all renormalization prescriptions.

According to \cite{Kazantsev:2020kfl}, the renormalization prescription giving the $\overline{\mbox{DR}}$ scheme for ${\cal N}=1$ supersymmetric theories with a single gauge coupling constant (regularized by higher covariant derivatives) is given by the equations

\begin{eqnarray}\label{Alpha_DR}
&&\hspace*{-5mm} \frac{1}{\alpha} -\frac{1}{\alpha_0} = -\frac{3}{2\pi} C_2 \Big(\ln\frac{\Lambda}{\mu} +\ln a_\varphi\Big) + \frac{1}{2\pi} T(R)\Big(\ln\frac{\Lambda}{\mu} +\ln a\Big)
- \frac{3\alpha}{4\pi^2} (C_2)^2 \Big(\ln\frac{\Lambda}{\mu} + \frac{1}{4} +\ln a_\varphi\Big) \nonumber\\
&&\hspace*{-5mm} + \frac{\alpha}{4\pi^2 r} C_2\, \mbox{tr}\, C(R)\, \Big(\ln\frac{\Lambda}{\mu} + \frac{1}{4} +\ln a \Big)
+ \frac{\alpha}{2\pi^2 r} \mbox{tr} \left(C(R)^2\right)\Big(\ln\frac{\Lambda}{\mu} -\frac{1}{4} -\frac{A}{2}\Big)
- \frac{1}{8\pi^3 r} C(R)_j{}^i \lambda^*_{imn}\nonumber\\
&&\hspace*{-5mm} \times \lambda^{jmn} \Big(\ln\frac{\Lambda}{\mu} - \frac{1}{4}-\frac{B}{2}\Big) + O(\alpha^2,\alpha\lambda^2,\lambda^4);\\
\label{LnZ_DR}
&&\hspace*{-5mm} \ln Z_i{}^j = \frac{\alpha}{\pi} C(R)_i{}^j\Big(\ln\frac{\Lambda}{\mu} - \frac{1}{2} - \frac{A}{2}\Big)
- \frac{1}{4\pi^2} \lambda^*_{imn} \lambda^{jmn} \Big(\ln\frac{\Lambda}{\mu} -\frac{1}{2} - \frac{B}{2}\Big) + O(\alpha^2,\alpha\lambda^2,\lambda^4),
\end{eqnarray}

\noindent
where $r$ denotes the dimension of the (simple) gauge group. Also in the $\overline{\mbox{DR}}$ scheme the equations (\ref{Usual_Yukawa_Renormalization_U}) --- (\ref{Usual_Yukawa_Renormalization_E}) are certainly satisfied. This implies that the constants $j$ can be obtained from Eqs. (\ref{J_Vs_G1}) --- (\ref{J_Vs_G3}).

The generalization of the relations (\ref{Alpha_DR}) and (\ref{LnZ_DR}) to the case of multiple gauge couplings can be written as

\begin{eqnarray}
&&\hspace*{-5mm} \frac{1}{\alpha_K} -\frac{1}{\alpha_{0K}} = -\frac{3}{2\pi} C_2(G_K) \Big(\ln\frac{\Lambda}{\mu} +\ln a_{\varphi,K}\Big) + \frac{1}{2\pi} \sum\limits_{\mbox{\scriptsize a}} \bm{T}_{\mbox{\scriptsize a}K}\Big(\ln\frac{\Lambda}{\mu} +\ln a_K\Big)
\nonumber\\
&&\hspace*{-5mm} - \frac{3\alpha_K}{4\pi^2} \left(C_2(G_K)\vphantom{{}^2}\right)^2 \Big(\ln\frac{\Lambda}{\mu} + \frac{1}{4} +\ln a_{\varphi,K}\Big) + \frac{\alpha_K}{4\pi^2} C_2(G_K) \sum\limits_{\mbox{\scriptsize a}} \bm{T}_{\mbox{\scriptsize a}K} \Big(\ln\frac{\Lambda}{\mu} + \frac{1}{4} +\ln a_K \Big)
\nonumber\\
&&\hspace*{-5mm} + \frac{1}{2\pi^2} \sum\limits_{L,\mbox{\scriptsize a}} \alpha_L \bm{T}_{\mbox{\scriptsize a}K} C(R_{\mbox{\scriptsize a}L}) \Big(\ln\frac{\Lambda}{\mu} -\frac{1}{4} -\frac{A}{2}\Big)
- \frac{1}{8\pi^3} \sum\limits_{\mbox{\scriptsize a}} \bm{T}_{\mbox{\scriptsize a}K} (\lambda^* \lambda)_{\mbox{\scriptsize a}}{}^{\mbox{\scriptsize a}} \Big(\ln\frac{\Lambda}{\mu} - \frac{1}{4}-\frac{B}{2}\Big)
\nonumber\\
&&\hspace*{-5mm} + O(\alpha^2,\alpha\lambda^2,\lambda^4);\vphantom{\frac{1}{2}}\\
&&\hspace*{-5mm} \ln Z_{\mbox{\scriptsize a}}{}^{\mbox{\scriptsize b}} = \sum\limits_K \frac{\alpha_K}{\pi} C(R_{\mbox{\scriptsize a}K}) \delta_{\mbox{\scriptsize a}}{}^{\mbox{\scriptsize b}} \Big(\ln\frac{\Lambda}{\mu} - \frac{1}{2} - \frac{A}{2}\Big)
- \frac{1}{4\pi^2} (\lambda^*\lambda)_{\mbox{\scriptsize a}}{}^{\mbox{\scriptsize b}} \Big(\ln\frac{\Lambda}{\mu} -\frac{1}{2} - \frac{B}{2}\Big)\nonumber\\
&&\hspace*{-5mm} + O(\alpha^2,\alpha\lambda^2,\lambda^4).
\end{eqnarray}

\noindent
Substituting $\bm{T}_{\mbox{\scriptsize a}K}$ and $(\lambda^*\lambda)_{\mbox{\scriptsize a}}{}^{\mbox{\scriptsize b}}$ from Tables \ref{Table_MSSM_TaK} and \ref{Table_MSSM_LambdaLambda}, respectively, and comparing the results with Eqs. (\ref{Alpha3}) --- (\ref{Z_Hd}) we obtain the values of the finite constants

\begin{eqnarray}\label{Finite_Constants_G_DR}
&& g_{\mbox{\scriptsize a}1} =  g_{\mbox{\scriptsize a}2} = g_{\mbox{\scriptsize a}3} = -\frac{1}{2}-\frac{A}{2}; \qquad\qquad\qquad\quad\ \, g_{\mbox{\scriptsize ab}} = -\frac{1}{2}-\frac{B}{2};\\
\label{Finite_Constants_B1_DR}
&& b_{1,1} = \ln a_1;\qquad b_{1,2} = 7\ln a_2 - 6\ln a_{\varphi,2};\qquad b_{1,3} = -2\ln a_3 + 3\ln a_{\varphi,3},\qquad\vphantom{\frac{1}{2}}\\
\label{Finite_Constants_B2_DR}
&& b_{2,11} = b_{2,12} = b_{2,13} = b_{2,21} = b_{2,23} = b_{2,31} = b_{2,32} = -\frac{1}{4} -\frac{A}{2};\nonumber\\
&& b_{2,22} = \frac{4}{25}\Big( - 6\ln a_{\varphi,2} + 7\ln a_2\Big)-\frac{21 A}{50} - \frac{17}{100};\qquad\nonumber\\
&& b_{2,33} = - \frac{9}{7}\Big(3\ln a_{\varphi,3} - 2\ln a_3 \Big) -\frac{8A}{7} - \frac{25}{28};\qquad \nonumber\\
&& b_{2,1\mbox{\scriptsize a}} = b_{2,2\mbox{\scriptsize a}} = b_{2,3\mbox{\scriptsize a}} = - \frac{1}{4} - \frac{B}{2},
\end{eqnarray}

\noindent
where $\mbox{a},\mbox{b} = Q,U,D,L,E,H_u,H_d$. (Certainly, $\mbox{a}$ and $\mbox{b}$ can take only such values for that the corresponding $g$ and $b$ are present in Eqs. (\ref{Alpha3}) --- (\ref{Alpha1}) or (\ref{Z_Q}) --- (\ref{Z_Hd}).) According to Eqs. (\ref{J_Vs_G1}) --- (\ref{J_Vs_G3}) and (\ref{Finite_Constants_G_DR}) the values of the constants $j$ in the $\overline{\mbox{DR}}$ scheme are

\begin{eqnarray}\label{Finite_Constants_J_DR}
&& j_{\mbox{\scriptsize a}1} =  j_{\mbox{\scriptsize a}2} = j_{\mbox{\scriptsize a}3} = -\frac{1}{2}-\frac{A}{2}; \qquad j_{\mbox{\scriptsize ab}} = -\frac{1}{2}-\frac{B}{2};\nonumber\\
&& j_{UtU} = j_{DtD} = j_{DtE} = j_{EtD} = j_{EtE} = -\frac{1}{2}-\frac{B}{2}.
\end{eqnarray}

\noindent
Substituting Eqs. (\ref{Finite_Constants_G_DR}) --- (\ref{Finite_Constants_J_DR}) into the two-loop anomalous dimensions (\ref{MSSM_Anomalous_Dimentions_Q_Renormalized}) --- (\ref{MSSM_Anomalous_Dimentions_Hd_Renormalized}) we obtain the expressions (\ref{MSSM_Anomalous_Dimentions_Q_DR}) --- (\ref{MSSM_Anomalous_Dimentions_Hd_DR}), which coincide with the result in the $\overline{\mbox{DR}}$ scheme, see, e.g., \cite{Jack:2004ch}. Similarly, substituting these values into the three-loop $\beta$-functions (\ref{Beta3_Renormalized}) --- (\ref{Beta1_Renormalized}) we reproduce the $\overline{\mbox{DR}}$ result (\ref{MSSM_Beta3_DR}) --- (\ref{MSSM_Beta1_DR}) obtained in \cite{Jack:2004ch}. Certainly, comparing the results it is necessary to take into account the difference in notation, namely,

\begin{eqnarray}
&& \alpha_{1} = \frac{g_{1}^2}{4\pi}; \qquad \alpha_{2} = \frac{g_{2}^2}{4\pi}; \qquad \alpha_{3} = \frac{g_{3}^2}{4\pi}; \qquad\, Y_U = Y_t;\qquad Y_D = Y_b;\qquad Y_E = Y_\tau; \nonumber\\
&& \widetilde\beta_1(\alpha,Y) = \frac{g_1}{2\pi} \beta_1(g,Y);\qquad \widetilde\beta_2(\alpha,Y) = \frac{g_2}{2\pi} \beta_2(g,Y);\qquad  \widetilde\beta_3(\alpha,Y) = \frac{g_3}{2\pi} \beta_3(g,Y);\qquad \nonumber\\
&& \widetilde\gamma_Q(\alpha,Y) = 2\gamma_Q(g,Y)^T;\qquad \widetilde\gamma_U(\alpha,Y) = 2\gamma_t(g,Y);\qquad\ \ \widetilde\gamma_D(\alpha,Y) = 2\gamma_b(g,Y);\qquad \vphantom{\frac{1}{2}}\nonumber\\
&& \widetilde\gamma_L(\alpha,Y) = 2\gamma_L(g,Y)^T;\qquad\, \widetilde\gamma_E(\alpha,Y) = 2\gamma_\tau(g,Y);\qquad\ \ \widetilde\gamma_{H_u}(\alpha,Y) = 2\gamma_{H_2}(g,Y);\qquad \vphantom{\frac{1}{2}}\nonumber\\
&& \widetilde\gamma_{H_d}(\alpha,Y) = 2\gamma_{H_1}(g,Y).\vphantom{\frac{1}{2}}
\end{eqnarray}

\section{A class of the NSVZ schemes for MSSM}
\hspace*{\parindent}\label{Section_NSVZ_Class}

Certainly, for RGFs defined in terms of the renormalized couplings the NSVZ equations

\begin{eqnarray}\label{MSSM_Exact_Beta3_Renormalized}
&&\hspace*{-5mm} \frac{\widetilde\beta_3(\alpha,Y)}{\alpha_{3}^2} = - \frac{1}{2\pi(1 - 3\alpha_{3}/2\pi)} \bigg[3 + \mbox{tr}\Big(\widetilde\gamma_{Q}(\alpha,Y) + \frac{1}{2} \widetilde\gamma_{U}(\alpha,Y) + \frac{1}{2} \widetilde\gamma_{D}(\alpha,Y)\Big)\bigg];\\
\label{MSSM_Exact_Beta2_Renormalized}
&&\hspace*{-5mm} \frac{\widetilde\beta_2(\alpha,Y)}{\alpha_{2}^2} = - \frac{1}{2\pi(1 - \alpha_{2}/\pi)} \bigg[-1 + \mbox{tr}\Big(\frac{3}{2} \widetilde\gamma_{Q}(\alpha,Y) + \frac{1}{2} \widetilde\gamma_{L}(\alpha,Y)\Big) + \frac{1}{2} \widetilde\gamma_{H_u}(\alpha,Y) \nonumber\\
&&\hspace*{-5mm} + \frac{1}{2} \widetilde\gamma_{H_d}(\alpha,Y)\bigg];\\
\label{MSSM_Exact_Beta1_Renormalized}
&&\hspace*{-5mm} \frac{\widetilde\beta_1(\alpha,Y)}{\alpha_{1}^2} = - \frac{3}{5} \cdot \frac{1}{2\pi}\bigg[-11 + \mbox{tr}\Big(\frac{1}{6} \widetilde\gamma_{Q}(\alpha,Y) + \frac{4}{3} \widetilde\gamma_{U}(\alpha,Y)
+ \frac{1}{3} \widetilde\gamma_{D}(\alpha,Y) + \frac{1}{2} \widetilde\gamma_{L}(\alpha,Y)  \nonumber\\
&&\hspace*{-5mm} + \widetilde\gamma_{E}(\alpha,Y)\Big) + \frac{1}{2} \widetilde\gamma_{H_u}(\alpha,Y) + \frac{1}{2} \widetilde\gamma_{H_d}(\alpha,Y)\bigg]
\end{eqnarray}

\noindent
are not satisfied for general renormalization prescriptions. However, comparing the anomalous dimensions  (\ref{MSSM_Anomalous_Dimentions_Q_Renormalized}) --- (\ref{MSSM_Anomalous_Dimentions_Hd_Renormalized}) and the gauge $\beta$-functions (\ref{Beta3_Renormalized}) --- (\ref{Beta1_Renormalized}) we see that the NSVZ equations for them are valid if the finite constants satisfy the equations

\begin{eqnarray}\label{NSVZ_Finite_Constants}
&&\hspace*{-5mm} b_{2,11} = \frac{1}{398}\Big(g_{Q1} +128 g_{U1} +8 g_{D1} +27 g_{L1} + 216 g_{E1} + 9 g_{H_u1} + 9 g_{H_d1}\Big);\qquad \nonumber\\
&&\hspace*{-5mm} b_{2,12} = \frac{1}{6}\Big(g_{Q2} + 3 g_{L2} + g_{H_u2} + g_{H_d2}\Big);\qquad\ \, b_{2,13} = \frac{1}{11}\Big(g_{Q3} + 8 g_{U3} + 2 g_{D3}\Big);\qquad \nonumber\\
&&\hspace*{-5mm} b_{2,21} = \frac{1}{6}\Big(g_{Q1} + 3 g_{L1} + g_{H_u1} + g_{H_d1}\Big);\qquad\ \,  b_{2,22} = \frac{1}{50}\Big(27 g_{Q2} + 9 g_{L2} + 3 g_{H_u2} +3g_{H_d2} + 8 b_{1,2}\Big); \nonumber\\
&&\hspace*{-5mm} b_{2,23} = g_{Q3};\qquad\quad\ b_{2,31} = \frac{1}{11}\Big(g_{Q1} + 8 g_{U1} + 2 g_{D1}\Big);\qquad\quad\ b_{2,32} = g_{Q2}; \nonumber\\
&&\hspace*{-5mm} b_{2,33} = \frac{1}{7}\Big(8 g_{Q3} + 4 g_{U3} + 4 g_{D3} - 9 b_{1,3}\Big);\qquad
b_{2,1U} = \frac{1}{26}\Big(g_{QU} + 16 g_{UU} + 9 g_{H_uU}\Big);\nonumber\\
&&\hspace*{-5mm} b_{2,1D} = \frac{1}{14}\Big(g_{QD} + 4 g_{DD} + 9 g_{H_dD}\Big);\qquad\qquad\  b_{2,1E} = \frac{1}{6}\Big(g_{LE} + 4 g_{EE} + g_{H_dE}\Big);\nonumber\\
&&\hspace*{-5mm} b_{2,2U} = \frac{1}{2}\Big(g_{QU} + g_{H_uU}\Big);\qquad b_{2,2D} = \frac{1}{2}\Big(g_{QD} + g_{H_dD}\Big);\qquad  b_{2,2E} = \frac{1}{2}\Big(g_{LE} + g_{H_dE}\Big);\nonumber\\
&&\hspace*{-5mm} b_{2,3U} = \frac{1}{2}\Big(g_{QU} + g_{UU}\Big);\qquad\ \, b_{2,3D} = \frac{1}{2}\Big(g_{QD} + g_{DD}\Big).
\end{eqnarray}

\noindent
(Note that the values of the constants $j$ are not restricted.) These equations can be compared with the general result obtained in \cite{Korneev:2021zdz}. Really, according to \cite{Korneev:2021zdz}, two different NSVZ schemes are related by the finite renormalization

\begin{equation}\label{Finite_Renormalization_General}
\alpha_K' = \alpha_K'(\alpha,\lambda);\qquad \lambda'=\lambda'(\alpha,\lambda);\qquad
Z_{\mbox{\scriptsize a}}'\Big(\alpha',\lambda',\ln\frac{\Lambda}{\mu}\Big) = z_{\mbox{\scriptsize a}}(\alpha,\lambda) Z_{\mbox{\scriptsize a}}\Big(\alpha,\lambda,\ln\frac{\Lambda}{\mu}\Big)
\end{equation}

\noindent
which satisfies the equation

\begin{equation}\label{NSVZ_Class}
\frac{1}{\alpha_K'} - \frac{1}{\alpha_K} + \frac{C_2(G_K)}{2\pi} \ln \frac{\alpha_K'}{\alpha_K} - \frac{1}{2\pi}\sum\limits_{\mbox{\scriptsize a}}\bm{T}_{\mbox{\scriptsize a}K} \ln z_{\mbox{\scriptsize a}} = B_K,
\end{equation}

\noindent
where $B_{K}$ are some constants. Eq. (\ref{NSVZ_Class}) generalizes similar results for theories with simple gauge groups derived in \cite{Goriachuk:2018cac,Goriachuk_Conference,Goriachuk:2020wyn}. For MSSM from Eq. (\ref{NSVZ_Class}) we obtain the constraints

\begin{eqnarray}\label{Alpha3_Equation}
&&\hspace*{-5mm}  \frac{1}{\alpha_3'} - \frac{1}{\alpha_3} + \frac{3}{2\pi} \ln \frac{\alpha_3'}{\alpha_3} - \frac{1}{2\pi}\mbox{tr}\Big(\ln (z_Q)^T + \frac{1}{2}\ln z_U + \frac{1}{2}\ln z_D\Big)=B_3;\\
\label{Alpha2_Equation}
&&\hspace*{-5mm}  \frac{1}{\alpha_2'} - \frac{1}{\alpha_2} + \frac{1}{\pi} \ln \frac{\alpha_2'}{\alpha_2} - \frac{1}{2\pi}\mbox{tr}\Big(\frac{3}{2}\ln (z_Q)^T + \frac{1}{2}\ln (z_L)^T \Big) - \frac{1}{2\pi}\Big(\frac{1}{2}\ln z_{H_u}
+ \frac{1}{2} \ln z_{H_d} \Big)=B_2;\qquad\\
\label{Alpha1_Equation}
&&\hspace*{-5mm} \frac{1}{\alpha_1'} - \frac{1}{\alpha_1} - \frac{1}{2\pi}\cdot \frac{3}{5}\, \mbox{tr}\Big(\frac{1}{6} \ln (z_Q)^T + \frac{4}{3}\ln z_U + \frac{1}{3}\ln z_D + \frac{1}{2}\ln (z_L)^T + \ln z_E\Big)\nonumber\\
&&\hspace*{-5mm} \qquad\qquad\qquad\qquad\qquad\qquad\qquad\qquad\qquad\qquad\quad\ \,
- \frac{1}{2\pi}\cdot \frac{3}{5}\Big(\frac{1}{2}\ln z_{H_u} + \frac{1}{2}\ln z_{H_d} \Big)=B_1.
\end{eqnarray}

\noindent
The finite renormalizations (\ref{Finite_Renormalization_General}) which relate the HD+MSL scheme to the scheme defined by Eqs. (\ref{Alpha3}) --- (\ref{Z_Hd}) are written in the form

\begin{eqnarray}\label{Finite_Renormalizations_Alpha}
&&\hspace*{-5mm} \frac{1}{\alpha'_{3}} =\frac{1}{\alpha_3} - \frac{1}{2\pi}\bigg[3 b_{1,3} -\frac{11\alpha_1}{20\pi} b_{2,31} -\frac{9\alpha_2}{4\pi} b_{2,32} - \frac{7\alpha_3}{2\pi}
b_{2,33}\nonumber\\
&&\hspace*{-5mm}\qquad\qquad\qquad\qquad\qquad\qquad\quad\, +\frac{1}{4\pi^2}\mbox{tr}\Big(Y_U^+ Y_U\Big) b_{2,3U} + \frac{1}{4\pi^2}\mbox{tr}\Big(Y_D^+ Y_D\Big) b_{2,3D}\bigg] + O(\alpha^2,\alpha Y^2, Y^4);\nonumber\\
&&\hspace*{-5mm} \frac{1}{\alpha'_{2}} =\frac{1}{\alpha_2} - \frac{1}{2\pi}\bigg[- b_{1,2} -\frac{9\alpha_1}{20\pi} b_{2,21} -\frac{25\alpha_2}{4\pi} b_{2,22} - \frac{6\alpha_3}{\pi} b_{2,23}
+\frac{3}{8\pi^2}\mbox{tr}\Big(Y_U^+ Y_U\Big) b_{2,2U} \nonumber\\
&&\hspace*{-5mm}\qquad\qquad\qquad\qquad\qquad\qquad\quad\,  + \frac{3}{8\pi^2}\mbox{tr}\Big(Y_D^+ Y_D\Big) b_{2,2D}
+ \frac{1}{8\pi^2}\mbox{tr}\Big(Y_E^+ Y_E\Big) b_{2,2E} \bigg] + O(\alpha^2,\alpha Y^2, Y^4);\nonumber\\
&&\hspace*{-5mm} \frac{1}{\alpha'_{1}} =\frac{1}{\alpha_1} - \frac{1}{2\pi}\cdot \frac{3}{5}\bigg[-11 b_{1,1} -\frac{199\alpha_1}{60\pi} b_{2,11} -\frac{9\alpha_2}{4\pi} b_{2,12} - \frac{22\alpha_3}{3\pi} b_{2,13}
+\frac{13}{24\pi^2}\mbox{tr}\Big(Y_U^+ Y_U\Big) b_{2,1U} \nonumber\\
&&\hspace*{-5mm} \qquad\qquad\qquad\qquad\qquad\qquad\ \  + \frac{7}{24\pi^2}\mbox{tr}\Big(Y_D^+ Y_D\Big) b_{2,1D}
+ \frac{3}{8\pi^2} \mbox{tr}\Big(Y_E^+ Y_E\Big) b_{2,1E} \bigg] + O(\alpha^2,\alpha Y^2, Y^4);\nonumber\\
&&\vphantom{1}\\
\label{Finite_Renormalizations_Z}
&&\hspace*{-5mm} (z_Q)^T = 1 + \frac{\alpha_1}{60\pi} g_{Q1} + \frac{3\alpha_2}{4\pi} g_{Q2} + \frac{4\alpha_3}{3\pi} g_{Q3} - \frac{1}{8\pi^2} Y_{U} Y_{U}^+ g_{QU} -\frac{1}{8\pi^2} Y_{D} Y_{D}^+ g_{QD}
+ O(\alpha^2,\alpha Y^2, Y^4);\nonumber\\
&&\hspace*{-5mm} z_U = 1 + \frac{4 \alpha_{1}}{15 \pi} g_{U1} + \frac{4\alpha_{3}}{3\pi} g_{U3} - \frac{1}{4\pi^2}\, Y_{U}^+ Y_{U}\, g_{UU} + O(\alpha^2,\alpha Y^2, Y^4);\nonumber\\
&&\hspace*{-5mm} z_D = 1 + \frac{\alpha_{1}}{15\pi} g_{D1} + \frac{4\alpha_{3}}{3\pi} g_{D3} - \frac{1}{4\pi^2}\, Y_{D}^+ Y_{D}\, g_{DD} + O(\alpha^2,\alpha Y^2, Y^4);\nonumber\\
&&\hspace*{-5mm} (z_L)^T = 1 + \frac{3\alpha_{1}}{20\pi} g_{L1} + \frac{3\alpha_{2}}{4\pi} g_{L2} - \frac{1}{8\pi^2}\, Y_{E} Y_{E}^+\, g_{LE} + O(\alpha^2,\alpha Y^2, Y^4);\nonumber\\
&&\hspace*{-5mm} z_E = 1 + \frac{3\alpha_{1}}{5\pi} g_{E1} - \frac{1}{4\pi^2}\, Y_{E}^+ Y_{E}\, g_{EE} + O(\alpha^2,\alpha Y^2, Y^4);\nonumber\\
&&\hspace*{-5mm} z_{H_u} = 1 + \frac{3\alpha_{1}}{20\pi} g_{H_u1} + \frac{3\alpha_{2}}{4\pi} g_{H_u2} - \frac{3}{8\pi^2}\, \mbox{tr}\Big(Y_{U}^+ Y_{U}\Big) g_{H_uU} + O(\alpha^2,\alpha Y^2, Y^4);\qquad\nonumber\\
&&\hspace*{-5mm} z_{H_d} = 1 + \frac{3\alpha_{1}}{20\pi} g_{H_d1} + \frac{3\alpha_{2}}{4\pi} g_{H_d2} - \frac{3}{8\pi^2}\, \mbox{tr}\Big( Y_{D}^+ Y_{D}\Big) g_{H_dD}
- \frac{1}{8\pi^2} \mbox{tr}\Big(Y_{E}^+ Y_{E}\Big) g_{H_dE} + O(\alpha^2,\alpha Y^2, Y^4).\nonumber\\
\end{eqnarray}

\noindent
(Here the couplings and renormalization constants without primes correspond to the HD+MSL scheme.) Substituting the expressions (\ref{Finite_Renormalizations_Alpha}) and (\ref{Finite_Renormalizations_Z}) into Eqs. (\ref{Alpha3_Equation}) --- (\ref{Alpha1_Equation}) we obtain the relations (\ref{NSVZ_Finite_Constants}) together with the equations

\begin{equation}
B_1 = \frac{33}{10\pi} b_{1,1};\qquad B_2 = \frac{1}{2\pi} b_{1,2};\qquad B_3 = - \frac{3}{2\pi} b_{1,3}.
\end{equation}

\noindent
Thus, Eq. (\ref{NSVZ_Class}) has been verified by a nontrivial explicit calculation in such an approximation where the scheme dependence becomes essential.

\section{Conclusion}
\hspace*{\parindent}

In this paper we found the two-loop anomalous dimensions for all MSSM chiral matter superfields in the case of using the higher covariant derivative regularization and for an arbitrary  renormalization prescription which supplements it. For this purpose we generalized the expression (\ref{Two_Loop_Gamma_Bare_Single_Gauge_Coupling}) obtained in \cite{Kazantsev:2020kfl} to theories with multiple gauge couplings and constructed the anomalous dimensions defined in terms of the bare couplings in the particular case of MSSM. They were considered as a starting point for calculating the anomalous dimensions standardly defined in terms of the renormalized couplings, which depend on some finite constants determining a subtraction scheme in the lowest approximation. Note that we use the manifestly ${\cal N}=1$ supersymmetric formulation of the theory in terms of ${\cal N}=1$ superfields and do not break supersymmetry introducing these finite constants. This implies that we consider a general renormalization prescription compatible with it.

It is known \cite{Stepanyantz:2020uke} that in ${\cal N}=1$ supersymmetric theories regularized by higher covariant derivatives RGFs defined in terms of the bare couplings satisfy the NSVZ equation in all orders independently of a renormalization prescription, at least, for theories with a single gauge coupling constant. (For a fixed regularization these RGFs simply do not depend on a renormalization prescription \cite{Kataev:2013eta}.) It is highly likely \cite{Korneev:2021zdz} that a similar statement is also valid for ${\cal N}=1$ supersymmetric theories with multiple gauge couplings. Therefore, it is possible to calculate the $\beta$-functions in a certain approximation starting from the anomalous dimensions in the previous orders. In particular, in this paper the NSVZ equations for theories with multiple gauge couplings \cite{Shifman:1996iy,Korneev:2021zdz} were applied for calculating the three-loop MSSM $\beta$-functions. Certainly, these equations should be applied for RGFs defined in terms of the bare couplings, because the standard RGFs satisfy them only for some special renormalization prescriptions, e.g., in the HD+MSL scheme \cite{Kataev:2013eta,Shakhmanov:2017wji,Stepanyantz:2017sqg}. Again, starting from the three-loop $\beta$-functions defined in terms of the bare couplings we obtained the (scheme-dependent) $\beta$-functions defined in terms of the renormalized couplings for an arbitrary supersymmetric renormalization prescription which is obtained if the renormalization constants for all components of each superfield coincide.

As a test of the calculation correctness, we checked that for a certain choice of a subtraction scheme the results (for both the two-loop anomalous dimensions and the three-loop $\beta$-functions) coincide with the ones obtained earlier in \cite{Jack:2004ch} in the $\overline{\mbox{DR}}$ scheme. Note that this test is not trivial due to the existence of some scheme dependent consequences of the NSVZ equations \cite{Kataev:2013csa,Kataev:2014gxa}.
Therefore, in this paper we also independently confirm the results of Refs. \cite{Bjorkman:1985mi,Jack:2004ch} for RGFs obtained in the $\overline{\mbox{DR}}$ scheme.

Moreover, we described the class of the NSVZ schemes for MSSM and demonstrated that the finite renormalizations relating different NSVZ schemes satisfy the equations (\ref{NSVZ_Class}) derived in \cite{Korneev:2021zdz}. Certainly the $\overline{\mbox{DR}}$ scheme does not enter this class according to \cite{Jack:1996vg,Jack:1996cn}.

\section*{Acknowledgments}
\hspace*{\parindent}

The work of K.S. was supported by Russian Scientific Foundation, grant No. 21-12-00129.

\appendix

\section*{Appendix}

\section{Two-loop anomalous dimensions of the matter superfields defined in terms of the bare couplings}
\hspace*{\parindent}\label{Appendix_Two_Loop_Gamma}

In this appendix we collect the two-loop anomalous dimensions of all MSSM chiral superfields defined in terms of the bare couplings in the case of using the higher covariant derivative regularization. They were obtained with the help of Eq. (\ref{Gamma_Multicharge}) and are given by the expressions

\begin{eqnarray}\label{MSSM_Anomalous_Dimentions_Q_Bare}
&&\hspace*{-5mm} \gamma_Q(\alpha_0,Y_0)^T = - \frac{\alpha_{01}}{60\pi} - \frac{3\alpha_{02}}{4\pi} - \frac{4\alpha_{03}}{3\pi} + \frac{1}{8\pi^2}\Big(Y_{0U} Y_{0U}^+ + Y_{0D} Y_{0D}^+\Big)
+\frac{1}{2\pi^2}\bigg[\frac{\alpha_{01}^2}{3600} + \frac{9\alpha_{02}^2}{16}  + \frac{16\alpha_{03}^2}{9} \nonumber\\
&&\hspace*{-5mm} +\frac{\alpha_{01}\alpha_{02}}{40} + \frac{2\alpha_{01}\alpha_{03}}{45} + 2\alpha_{02}\alpha_{03} -\frac{9\alpha_{02}^2}{2}\Big(\ln a_{\varphi,2}+1+\frac{A}{2}\Big) - 12\alpha_{03}^2\Big(\ln a_{\varphi,3}+1+\frac{A}{2}\Big)
\nonumber\\
&&\hspace*{-5mm} + \frac{11\alpha_{01}^2}{100} \Big(\ln a_{1}+1+\frac{A}{2}\Big) + \frac{21\alpha_{02}^2}{4}\Big(\ln a_{2}+1+\frac{A}{2}\Big) + 8\alpha_{03}^2\Big(\ln a_{3}+1+\frac{A}{2}\Big)\bigg]
+\frac{1}{8\pi^3} Y_{0U} Y_{0U}^+ \nonumber\\
&&\hspace*{-5mm} \times \bigg[\frac{\alpha_{01}}{5} + (B-A)\Big(\frac{13\alpha_{01}}{60} +\frac{3\alpha_{02}}{4} + \frac{4\alpha_{03}}{3} \Big)\bigg]
+ \frac{1}{8\pi^3} Y_{0D} Y_{0D}^+ \bigg[\frac{\alpha_{01}}{10} + (B-A)\Big(\frac{7\alpha_{01}}{60} +\frac{3\alpha_{02}}{4}\nonumber\\
&&\hspace*{-5mm} + \frac{4\alpha_{03}}{3} \Big) \bigg] -\frac{1}{(8\pi^2)^2}\bigg[(Y_{0U} Y_{0U}^+)^2 + (Y_{0D} Y_{0D}^+)^2 + \frac{3}{2} Y_{0U} Y_{0U}^+\, \mbox{tr}\Big(Y_{0U} Y_{0U}^+\Big) + \frac{3}{2} Y_{0D} Y_{0D}^+\, \mbox{tr}\Big(Y_{0D} Y_{0D}^+\Big)\nonumber\\
&&\hspace*{-5mm} + \frac{1}{2} Y_{0D} Y_{0D}^+\, \mbox{tr}\Big(Y_{0E} Y_{0E}^+\Big) \bigg] + O\Big(\alpha_0^3,\alpha_0^2 Y_0^2, \alpha_0 Y_0^4, Y_0^6\Big);\qquad\\
&&\hspace*{-5mm} \gamma_U(\alpha_0,Y_0) = - \frac{4 \alpha_{01}}{15 \pi} - \frac{4\alpha_{03}}{3\pi} + \frac{1}{4\pi^2}\, Y_{0U}^+ Y_{0U}
+ \frac{1}{2\pi^2}\bigg[\frac{16\alpha_{01}^2}{225} + \frac{16\alpha_{03}^2}{9} + \frac{32\alpha_{01}\alpha_{03}}{45} - 12 \alpha_{03}^2 \nonumber\\
&&\hspace*{-5mm} \times \Big(\ln a_{\varphi,3}+1+\frac{A}{2}\Big) + \frac{44\alpha_{01}^2}{25}\Big(\ln a_{1}+1+\frac{A}{2}\Big) + 8\alpha_{03}^2\Big(\ln a_{3}+1+\frac{A}{2}\Big)\bigg] + \frac{1}{8\pi^3} Y_{0U}^+ Y_{0U} \nonumber\\
&&\hspace*{-5mm} \times\bigg[-\frac{\alpha_{01}}{10} + \frac{3\alpha_{02}}{2} + (B-A)\Big(\frac{13\alpha_{01}}{30} + \frac{3\alpha_{02}}{2} + \frac{8\alpha_{03}}{3}\Big)\bigg] - \frac{1}{(8\pi^2)^2}\bigg[(Y_{0U}^+ Y_{0U})^2 + 3 Y_{0U}^+Y_{0U} \nonumber\\
&&\hspace*{-5mm} \times\, \mbox{tr}\Big(Y_{0U} Y_{0U}^+\Big) + Y_{0U}^+ Y_{0D} Y_{0D}^+ Y_{0U} \bigg]
+ O\Big(\alpha_0^3,\alpha_0^2 Y_0^2,\alpha_0 Y_0^4, Y_0^6\Big);\\
&&\hspace*{-5mm} \gamma_D(\alpha_0,Y_0) = - \frac{\alpha_{01}}{15\pi} - \frac{4\alpha_{03}}{3\pi} + \frac{1}{4\pi^2}\, Y_{0D}^+ Y_{0D}
+ \frac{1}{2\pi^2}\bigg[\frac{\alpha_{01}^2}{225} + \frac{16\alpha_{03}^2}{9} + \frac{8\alpha_{01}\alpha_{03}}{45} - 12 \alpha_{03}^2 \nonumber\\
&&\hspace*{-5mm} \times \Big(\ln a_{\varphi,3}+1+\frac{A}{2}\Big) + \frac{11\alpha_{01}^2}{25}\Big(\ln a_{1}+1+\frac{A}{2}\Big) + 8\alpha_{03}^2\Big(\ln a_{3}+1+\frac{A}{2}\Big)\bigg] + \frac{1}{8\pi^3} Y_{0D}^+ Y_{0D}
\nonumber\\
&&\hspace*{-5mm} \times\bigg[\frac{\alpha_{01}}{10} + \frac{3\alpha_{02}}{2} + (B-A)\Big(\frac{7\alpha_{01}}{30} + \frac{3\alpha_{02}}{2} + \frac{8\alpha_{03}}{3}\Big)\bigg] - \frac{1}{(8\pi^2)^2}\bigg[(Y_{0D}^+ Y_{0D})^2
+  Y_{0D}^+Y_{0D} \nonumber\\
&&\hspace*{-5mm}
\times\, \mbox{tr}\Big(3 Y_{0D} Y_{0D}^+ + Y_{0E} Y_{0E}^+\Big) + Y_{0D}^+ Y_{0U} Y_{0U}^+ Y_{0D} \bigg]
+ O\Big(\alpha_0^3,\alpha_0^2 Y_0^2,\alpha_0 Y_0^4, Y_0^6\Big);\\
&&\hspace*{-5mm} \gamma_L(\alpha_0,Y_0)^T = - \frac{3\alpha_{01}}{20\pi} - \frac{3\alpha_{02}}{4\pi} + \frac{1}{8\pi^2}\, Y_{0E} Y_{0E}^+
+ \frac{1}{2\pi^2}\bigg[ \frac{9\alpha_{01}^2}{400} + \frac{9\alpha_{02}^2}{16} + \frac{9\alpha_{01}\alpha_{02}}{40} -\frac{9\alpha_{02}^2}{2} \Big(\ln a_{\varphi, 2}  \nonumber\\
&&\hspace*{-5mm} + 1 +\frac{A}{2} \Big) + \frac{99\alpha_{01}^2}{100} \Big(\ln a_{1} + 1+\frac{A}{2} \Big)
+ \frac{21\alpha_{02}^2}{4} \Big(\ln a_{2} + 1 +\frac{A}{2} \Big)\bigg] + \frac{1}{8\pi^3} Y_{0E} Y_{0E}^+ \bigg[\frac{3\alpha_{01}}{10}\nonumber\\
&&\hspace*{-5mm} + (B-A)\Big(\frac{9\alpha_{01}}{20} + \frac{3\alpha_{02}}{4}\Big)\bigg]
- \frac{1}{(8\pi^2)^2}\bigg[(Y_{0E} Y_{0E}^+)^2 + Y_{0E} Y_{0E}^+\, \mbox{tr} \Big(\frac{3}{2} Y_{0D}^+ Y_{0D} + \frac{1}{2} Y_{0E}^+ Y_{0E}\Big)\bigg]\nonumber\\
&&\hspace*{-5mm}
+ O\Big(\alpha_0^3,\alpha_0^2 Y_0^2,\alpha_0 Y_0^4, Y_0^6\Big);\\
&&\hspace*{-5mm} \gamma_E(\alpha_0,Y_0) = - \frac{3\alpha_{01}}{5\pi} + \frac{1}{4\pi^2}\, Y_{0E}^+ Y_{0E} + \frac{1}{2\pi^2}\cdot \frac{9\alpha_{01}^2}{25} \bigg[1+11 \Big(\ln a_1+1+\frac{A}{2}\Big)\bigg] + \frac{1}{8\pi^3} Y_{0E}^+ Y_{0E} \nonumber\\
&&\hspace*{-5mm} \times \bigg[-\frac{3\alpha_{01}}{10} + \frac{3\alpha_{02}}{2} + (B-A)\Big(\frac{9\alpha_{01}}{10} + \frac{3\alpha_{02}}{2}\Big)\bigg]
- \frac{1}{(8\pi^2)^2}\bigg[(Y_{0E}^+ Y_{0E})^2 + Y_{0E}^+ Y_{0E}\, \mbox{tr} \Big(3 Y_{0D}^+ Y_{0D}\nonumber\\
&&\hspace*{-5mm} + Y_{0E}^+ Y_{0E}\Big)\bigg] + O\Big(\alpha_0^3,\alpha_0^2 Y_0^2,\alpha_0 Y_0^4, Y_0^6\Big);\\
&&\hspace*{-5mm} \gamma_{H_u}(\alpha_0,Y_0) = - \frac{3\alpha_{01}}{20\pi} - \frac{3\alpha_{02}}{4\pi} + \frac{3}{8\pi^2}\, \mbox{tr}\Big(Y_{0U}^+ Y_{0U}\Big) + \frac{1}{2\pi^2}\bigg[ \frac{9\alpha_{01}^2}{400} + \frac{9\alpha_{02}^2}{16} + \frac{9\alpha_{01}\alpha_{02}}{40} -\frac{9\alpha_{02}^2}{2}\nonumber\\
&&\hspace*{-5mm} \times \Big(\ln a_{\varphi, 2} + 1+\frac{A}{2} \Big) + \frac{99\alpha_{01}^2}{100} \Big(\ln a_{1} + 1+\frac{A}{2} \Big) + \frac{21\alpha_{02}^2}{4} \Big(\ln a_{2} + 1+\frac{A}{2} \Big)\bigg]
+\frac{1}{8\pi^3} \mbox{tr}\Big(Y_{0U} Y_{0U}^+\Big)\nonumber\\
&&\hspace*{-5mm} \times \bigg[\frac{\alpha_{01}}{5} + 4\alpha_{03} + (B-A)\Big(\frac{13\alpha_{01}}{20} + \frac{9\alpha_{02}}{4} + 4\alpha_{03}\Big)\bigg]
-\frac{1}{(8\pi^2)^2}\bigg[\frac{3}{2}\,\mbox{tr}\Big(Y_{0D} Y_{0D}^+ Y_{0U} Y_{0U}^+\Big)\nonumber\\
&&\hspace*{-5mm} + \frac{9}{2}\,\mbox{tr}\Big((Y_{0U}Y_{0U}^+)^2\Big)\bigg]
+ O\Big(\alpha_0^3,\alpha_0^2 Y_0^2,\alpha_0 Y_0^4, Y_0^6\Big);\\
\label{MSSM_Anomalous_Dimentions_Hd_Bare}
&&\hspace*{-5mm} \gamma_{H_d}(\alpha_0,Y_0) = - \frac{3\alpha_{01}}{20\pi} - \frac{3\alpha_{02}}{4\pi} + \frac{1}{8\pi^2}\, \mbox{tr}\Big(3\, Y_{0D}^+ Y_{0D} + Y_{0E}^+ Y_{0E}\Big)  + \frac{1}{2\pi^2}\bigg[ \frac{9\alpha_{01}^2}{400} + \frac{9\alpha_{02}^2}{16} + \frac{9}{40} \nonumber\\
&&\hspace*{-5mm} \times \alpha_{01}\alpha_{02} -\frac{9\alpha_{02}^2}{2}\Big(\ln a_{\varphi, 2} + 1+\frac{A}{2} \Big) + \frac{99\alpha_{01}^2}{100} \Big(\ln a_{1} + 1+\frac{A}{2} \Big)
+ \frac{21\alpha_{02}^2}{4} \Big(\ln a_{2} + 1 +\frac{A}{2} \Big)\bigg]\nonumber\\
&&\hspace*{-5mm}  + \frac{1}{8\pi^3}\mbox{tr}\Big(Y_{0E} Y_{0E}^+\Big)\bigg[\frac{3\alpha_{01}}{10} + (B-A)\Big(\frac{9\alpha_{01}}{20} + \frac{3\alpha_{02}}{4}\Big)\bigg] + \frac{1}{8\pi^3} \mbox{tr}\Big(Y_{0D} Y_{0D}^+\Big)\bigg[-\frac{\alpha_{01}}{10} + 4\alpha_{03}\nonumber\\
&&\hspace*{-5mm} + (B-A)\Big(\frac{7\alpha_{01}}{20} + \frac{9\alpha_{02}}{4} + 4\alpha_{03}\Big)\bigg] - \frac{1}{(8\pi^2)^2}\bigg[\frac{3}{2}\mbox{tr}\Big((Y_{0E} Y_{0E}^+)^2\Big) + \frac{3}{2}\mbox{tr}\Big(Y_{0D} Y_{0D}^+ Y_{0U} Y_{0U}^+\Big)\nonumber\\
&&\hspace*{-5mm} + \frac{9}{2} \mbox{tr}\Big((Y_{0D} Y_{0D}^+)^2\Big)\bigg]
+ O\Big(\alpha_0^3,\alpha_0^2 Y_0^2,\alpha_0 Y_0^4, Y_0^6\Big).
\end{eqnarray}

\section{Three-loop MSSM $\beta$-functions defined in terms of the bare couplings}
\hspace{\parindent}\label{Appendix_Three_Loop_Beta}

Below we present the three-loop expressions for three MSSM gauge $\beta$-functions defined in terms of the bare couplings. They are written for the case of using the higher covariant derivative regularization and were obtained from the NSVZ equations (\ref{MSSM_Exact_Beta3}) --- (\ref{MSSM_Exact_Beta1}) in which we had substituted the two-loop expressions for the anomalous dimensions (\ref{MSSM_Anomalous_Dimentions_Q_Bare}) --- (\ref{MSSM_Anomalous_Dimentions_Hd_Bare}). The result has the form

\begin{eqnarray}
&&\hspace*{-5mm} \frac{\beta_3(\alpha_0,Y_0)}{\alpha_{03}^2} = - \frac{1}{2\pi} \bigg\{3 -\frac{11\alpha_{01}}{20\pi} -\frac{9\alpha_{02}}{4\pi} -\frac{7\alpha_{03}}{2\pi}
+ \frac{1}{8\pi^2}\, \mbox{tr}\Big(2\, Y_{0U}^+ Y_{0U} + 2\, Y_{0D}^+ Y_{0D}\Big)\nonumber\\
&&\hspace*{-5mm} + \frac{1}{2\pi^2}\bigg[ \frac{137\alpha_{01}^2}{1200} + \frac{27\alpha_{02}^2}{16} + \frac{\alpha_{03}^2}{6} + \frac{3\alpha_{01}\alpha_{02}}{40} - \frac{11\alpha_{01}\alpha_{03}}{60} -\frac{3\alpha_{02}\alpha_{03}}{4} - \frac{27\alpha_{02}^2}{2}\Big(\ln a_{\varphi,2} + 1 \nonumber\\
&&\hspace*{-5mm} +\frac{A}{2}\Big) - 72\alpha_{03}^2 \Big(\ln a_{\varphi,3} + 1 +\frac{A}{2}\Big) + \frac{363\alpha_{01}^2}{100}\Big(\ln a_{1} + 1 +\frac{A}{2}\Big) + \frac{63\alpha_{02}^2}{4}\Big(\ln a_{2} + 1 +\frac{A}{2}\Big) \nonumber\\
&&\hspace*{-5mm} + 48\alpha_{03}^2 \Big(\ln a_{3} + 1 +\frac{A}{2}\Big)\bigg]
+ \frac{1}{8\pi^3}\mbox{tr}\Big(Y_{0U} Y_{0U}^+\Big)\bigg[\frac{3\alpha_{01}}{20} + \frac{3\alpha_{02}}{4} + 3\alpha_{03} + (B-A)\Big(\frac{13\alpha_{01}}{30}
\nonumber\\
&&\hspace*{-5mm} + \frac{3\alpha_{02}}{2} + \frac{8\alpha_{03}}{3}\Big)\bigg] + \frac{1}{8\pi^3}\mbox{tr}\Big(Y_{0D} Y_{0D}^+\Big)\bigg[\frac{3\alpha_{01}}{20} + \frac{3\alpha_{02}}{4} + 3\alpha_{03} + (B-A)\Big(\frac{7\alpha_{01}}{30} + \frac{3\alpha_{02}}{2} \nonumber\\
&&\hspace*{-5mm} + \frac{8\alpha_{03}}{3}\Big)\bigg] - \frac{1}{(8\pi^2)^2}\bigg[\, \frac{3}{2}\mbox{tr}\Big((Y_{0U} Y_{0U}^+)^2 + (Y_{0D} Y_{0D}^+)^2\Big) + 3\Big(\mbox{tr}(Y_{0U} Y_{0U}^+)\Big)^2 + 3\Big(\mbox{tr}(Y_{0D} Y_{0D}^+)\Big)^2\nonumber\\
&&\hspace*{-5mm} + \mbox{tr}\Big(Y_{0E} Y_{0E}^+\Big)\,\mbox{tr}\Big(Y_{0D} Y_{0D}^+\Big) + \mbox{tr}\Big(Y_{0D} Y_{0D}^+ Y_{0U} Y_{0U}^+\Big)\bigg]\bigg\}
+ O(\alpha_0^3,\alpha_0^2 Y_0^2, \alpha_0 Y_0^4, Y_0^6);\qquad\\
&&\vphantom{1}\nonumber\\
&&\hspace*{-5mm} \frac{\beta_2(\alpha_0,Y_0)}{\alpha_{02}^2} = - \frac{1}{2\pi} \bigg\{-1  -\frac{9\alpha_{01}}{20\pi}- \frac{25\alpha_{02}}{4\pi} -\frac{6\alpha_{03}}{\pi} + \frac{1}{8\pi^2}\, \mbox{tr}\Big(3\, Y_{0U}^+ Y_{0U} + 3\, Y_{0D}^+ Y_{0D} + Y_{0E}^+ Y_{0E} \Big)\nonumber\\
&&\hspace*{-5mm} + \frac{1}{2\pi^2}\bigg[\frac{23\alpha_{01}^2}{400} - \frac{137\alpha_{02}^2}{16} + 8 \alpha_{03}^2 - \frac{9\alpha_{01}\alpha_{02}}{40} + \frac{\alpha_{01}\alpha_{03}}{5} - 3\alpha_{02}\alpha_{03}
- \frac{63\alpha_{02}^2}{2}\Big(\ln a_{\varphi,2} + 1 + \frac{A}{2}\Big)\nonumber\\
&&\hspace*{-5mm} -54\alpha_{03}^2\Big(\ln a_{\varphi,3} + 1 + \frac{A}{2}\Big) + \frac{297\alpha_{01}^2}{100}\Big(\ln a_{1} + 1 + \frac{A}{2}\Big) + \frac{147\alpha_{02}^2}{4}\Big(\ln a_{2} + 1 + \frac{A}{2}\Big) + 36\alpha_{03}^2\nonumber\\
&&\hspace*{-5mm} \times \Big(\ln a_{3} + 1 + \frac{A}{2}\Big)\bigg]
+ \frac{1}{8\pi^3}\mbox{tr}\Big(Y_{0U} Y_{0U}^+\Big)\bigg[\,\frac{2\alpha_{01}}{5} + 3\alpha_{02} + 2\alpha_{03} + (B-A)\Big(\frac{13\alpha_{01}}{20} + \frac{9\alpha_{02}}{4}\nonumber\\
&&\hspace*{-5mm} + 4\alpha_{03}\Big)\bigg]
+ \frac{1}{8\pi^3}\mbox{tr}\Big(Y_{0D} Y_{0D}^+\Big)\bigg[\,\frac{\alpha_{01}}{10}+3\alpha_{02}+2\alpha_{03}+(B-A)\Big(\frac{7\alpha_{01}}{20} + \frac{9\alpha_{02}}{4} + 4\alpha_{03}\Big)\bigg]\nonumber\\
&&\hspace*{-5mm} + \frac{1}{8\pi^3}\mbox{tr}\Big(Y_{0E} Y_{0E}^+\Big)\bigg[\,\frac{3\alpha_{01}}{10} + \alpha_{02} + (B-A)\Big(\frac{9\alpha_{01}}{20} + \frac{3\alpha_{02}}{4}\Big)\bigg]
- \frac{1}{(8\pi^2)^2}\bigg[\, \frac{5}{4}\mbox{tr}\Big(3(Y_{0U} Y_{0U}^+)^2\nonumber\\
&&\hspace*{-5mm} + 3(Y_{0D} Y_{0D}^+)^2 + (Y_{0E} Y_{0E}^+)^2\Big) + \frac{9}{4}\Big(\mbox{tr}(Y_{0U} Y_{0U}^+)\Big)^2 + \frac{9}{4}\Big(\mbox{tr}(Y_{0D} Y_{0D}^+)\Big)^2  + \frac{1}{4}\Big(\mbox{tr}(Y_{0E} Y_{0E}^+)\Big)^2
\nonumber\\
&&\hspace*{-5mm} + \frac{3}{2} \mbox{tr}\Big(Y_{0E} Y_{0E}^+\Big)\,\mbox{tr}\Big(Y_{0D} Y_{0D}^+\Big) + \frac{3}{2}\mbox{tr}\Big(Y_{0D} Y_{0D}^+ Y_{0U} Y_{0U}^+\Big) \bigg] \bigg\}
+ O(\alpha_0^3,\alpha_0^2 Y_0^2, \alpha_0 Y_0^4, Y_0^6);\\
&&\vphantom{1}\nonumber\\
&&\hspace*{-5mm} \frac{\beta_1(\alpha_0,Y_0)}{\alpha_{01}^2} = - \frac{1}{2\pi}\cdot \frac{3}{5} \bigg\{ -11 -\frac{199\alpha_{01}}{60\pi} -\frac{9\alpha_{02}}{4\pi} -\frac{22\alpha_{03}}{3\pi}
+ \frac{1}{8\pi^2} \mbox{tr}\Big(\frac{13}{3} Y_{0U}^+ Y_{0U} + \frac{7}{3} Y_{0D}^+ Y_{0D}\nonumber\\
&&\hspace*{-5mm} + 3 Y_{0E}^+ Y_{0E} \Big)
+ \frac{1}{2\pi^2}\bigg[\, \frac{5131\alpha_{01}^2}{3600} + \frac{27\alpha_{02}^2}{16} + \frac{88\alpha_{03}^2}{9} + \frac{23\alpha_{01}\alpha_{02}}{40} + \frac{137\alpha_{01}\alpha_{03}}{45} + \alpha_{02}\alpha_{03}
 \nonumber\\
&&\hspace*{-5mm} - \frac{27\alpha_{02}^2}{2}\Big(\ln a_{\varphi,2}+1+\frac{A}{2}\Big) - 66 \alpha_{03}^2 \Big(\ln a_{\varphi,3}+1+\frac{A}{2}\Big) + \frac{2189\alpha_{01}^2}{100}\Big(\ln a_{1}+1+\frac{A}{2}\Big)
\nonumber\\
&&\hspace*{-5mm} + \frac{63\alpha_{02}^2}{4}\Big(\ln a_{2}+1+\frac{A}{2}\Big) + 44\alpha_{03}^2 \Big(\ln a_{3}+1+\frac{A}{2}\Big) \bigg]
+ \frac{1}{8\pi^3}\mbox{tr}\Big(Y_{0U} Y_{0U}^+\Big)\bigg[\,2\alpha_{02} + 2\alpha_{03}\nonumber\\
&&\hspace*{-5mm} + (B-A)\Big(\frac{169\alpha_{01}}{180} + \frac{13\alpha_{02}}{4} + \frac{52\alpha_{03}}{9} \Big)\bigg]
+ \frac{1}{8\pi^3}\mbox{tr}\Big(Y_{0D} Y_{0D}^+\Big)\bigg[\,\frac{\alpha_{02}}{2} + 2\alpha_{03} + (B-A)\Big(\frac{49\alpha_{01}}{180}\nonumber\\
&&\hspace*{-5mm} + \frac{7\alpha_{02}}{4} + \frac{28\alpha_{03}}{9} \Big)\bigg]
+ \frac{1}{8\pi^3}\mbox{tr}\Big(Y_{0E} Y_{0E}^+\Big)\bigg[\,\frac{3\alpha_{02}}{2} +(B-A)\Big(\frac{27\alpha_{01}}{20} + \frac{9\alpha_{02}}{4} \Big)\bigg]
- \frac{1}{(8\pi^2)^2}\nonumber\\
&&\hspace*{-5mm} \times \bigg[\,\frac{15}{4} \mbox{tr}\Big((Y_{0U} Y_{0U}^+)^2\Big) + \frac{11}{4} \mbox{tr} \Big((Y_{0D} Y_{0D}^+)^2\Big) + \frac{19}{6}\mbox{tr}\Big(Y_{0D} Y_{0D}^+ Y_{0U} Y_{0U}^+\Big) + \frac{9}{4} \mbox{tr}\Big((Y_{0E} Y_{0E}^+)^2\Big)
\nonumber\\
&&\hspace*{-5mm} + \frac{17}{4} \Big(\mbox{tr}(Y_{0U} Y_{0U}^+)\Big)^2 + \frac{5}{4} \Big(\mbox{tr}(Y_{0D} Y_{0D}^+)\Big)^2 + \frac{5}{4} \Big(\mbox{tr}(Y_{0E} Y_{0E}^+)\Big)^2
+ \frac{25}{6} \mbox{tr}\Big(Y_{0E} Y_{0E}^+\Big)\,\mbox{tr}\Big(Y_{0D} Y_{0D}^+\Big)  \bigg]\bigg\}\nonumber\\
&&\hspace*{-5mm} + O(\alpha_0^3,\alpha_0^2 Y_0^2, \alpha_0 Y_0^4, Y_0^6).
\end{eqnarray}

\section{RGFs for MSSM in the $\overline{\mbox{DR}}$ scheme}
\hspace*{\parindent}\label{Appendix_RGFs_DR}

For completeness, in this appendix we present the expressions for MSSM RGFs in the $\overline{\mbox{DR}}$ scheme. Namely, the two-loop anomalous dimensions of the chiral matter superfields in our notation are written as

\begin{eqnarray}\label{MSSM_Anomalous_Dimentions_Q_DR}
&&\hspace*{-7mm} \widetilde\gamma_Q(\alpha,Y)^T = - \frac{\alpha_1}{60\pi} - \frac{3\alpha_2}{4\pi} - \frac{4\alpha_3}{3\pi} + \frac{1}{8\pi^2}\Big(Y_{U} Y_{U}^+ + Y_{D} Y_{D}^+\Big)
+\frac{1}{2\pi^2}\Big(\frac{199\alpha_{1}^2}{3600} + \frac{15\alpha_{2}^2}{16} - \frac{2\alpha_{3}^2}{9} \nonumber\\
&&\hspace*{-7mm}  +\frac{\alpha_{1}\alpha_{2}}{40} + \frac{2\alpha_{1}\alpha_{3}}{45} + 2\alpha_{2}\alpha_{3}\Big)
+\frac{1}{8\pi^3} Y_{U} Y_{U}^+ \cdot \frac{\alpha_{1}}{5} + \frac{1}{8\pi^3} Y_{D} Y_{D}^+ \cdot\frac{\alpha_{1}}{10} -\frac{1}{(8\pi^2)^2} \bigg[(Y_{U} Y_{U}^+)^2 + (Y_{D} Y_{D}^+)^2\nonumber\\
&&\hspace*{-7mm}  + \frac{3}{2} Y_{U} Y_{U}^+\, \mbox{tr}\Big(Y_{U} Y_{U}^+\Big) + \frac{3}{2} Y_{D} Y_{D}^+\, \mbox{tr}\Big(Y_{D} Y_{D}^+\Big)
+ \frac{1}{2} Y_{D} Y_{D}^+\, \mbox{tr}\Big(Y_{E} Y_{E}^+\Big) \bigg] + O\Big(\alpha^3,\alpha^2 Y^2, \alpha Y^4, Y^6\Big);\nonumber\\
&&\vphantom{1}\\
&&\hspace*{-7mm} \widetilde\gamma_U(\alpha,Y) = - \frac{4 \alpha_{1}}{15 \pi} - \frac{4\alpha_{3}}{3\pi} + \frac{1}{4\pi^2}\, Y_{U}^+ Y_{U}
+ \frac{1}{2\pi^2}\Big(\frac{214\alpha_{1}^2}{225} - \frac{2\alpha_{3}^2}{9} + \frac{32\alpha_{1}\alpha_{3}}{45} \Big)
+ \frac{1}{8\pi^3} Y_{U}^+ Y_{U} \Big(-\frac{\alpha_{1}}{10} \nonumber\\
&&\hspace*{-7mm} + \frac{3\alpha_{2}}{2}\Big)  - \frac{1}{(8\pi^2)^2}\bigg[(Y_{U}^+ Y_{U})^2 + 3 Y_{U}^+Y_{U}\, \mbox{tr}\Big(Y_{U} Y_{U}^+\Big) + Y_{U}^+ Y_{D} Y_{D}^+ Y_{U} \bigg]
+ O\Big(\alpha^3,\alpha^2 Y^2,\alpha Y^4, Y^6\Big);\nonumber\\
&&\vphantom{1}\\
&&\hspace*{-7mm} \widetilde\gamma_D(\alpha,Y) = - \frac{\alpha_{1}}{15\pi} - \frac{4\alpha_{3}}{3\pi} + \frac{1}{4\pi^2}\, Y_{D}^+ Y_{D}
+ \frac{1}{2\pi^2}\Big( \frac{101\alpha_{1}^2}{450} - \frac{2\alpha_{3}^2}{9} + \frac{8\alpha_{1}\alpha_{3}}{45}\Big) + \frac{1}{8\pi^3}
\nonumber\\
&&\hspace*{-7mm} \times Y_{D}^+ Y_{D} \Big(\frac{\alpha_{1}}{10} + \frac{3\alpha_{2}}{2}\Big) - \frac{1}{(8\pi^2)^2}\bigg[(Y_{D}^+ Y_{D})^2
+  Y_{D}^+Y_{D}\, \mbox{tr}\Big(3 Y_{D} Y_{D}^+ + Y_{E} Y_{E}^+\Big) + Y_{D}^+ Y_{U} Y_{U}^+ Y_{D} \bigg]\nonumber\\
&&\hspace*{-7mm}
+ O\Big(\alpha^3,\alpha^2 Y^2,\alpha Y^4, Y^6\Big);\\
&&\hspace*{-7mm} \widetilde\gamma_L(\alpha,Y)^T = - \frac{3\alpha_{1}}{20\pi} - \frac{3\alpha_{2}}{4\pi} + \frac{1}{8\pi^2}\, Y_{E} Y_{E}^+
+ \frac{1}{2\pi^2}\Big(\frac{207\alpha_{1}^2}{400} + \frac{15\alpha_{2}^2}{16} + \frac{9\alpha_{1}\alpha_{2}}{40}\Big) + \frac{1}{8\pi^3} Y_{E} Y_{E}^+ \cdot \frac{3\alpha_{1}}{10}\nonumber\\
&&\hspace*{-7mm}
- \frac{1}{(8\pi^2)^2}\bigg[(Y_{E} Y_{E}^+)^2 + Y_{E} Y_{E}^+\, \mbox{tr} \Big(\frac{3}{2} Y_{D}^+ Y_{D} + \frac{1}{2} Y_{E}^+ Y_{E}\Big)\bigg]
+ O\Big(\alpha^3,\alpha^2 Y^2,\alpha Y^4, Y^6\Big);\\
&&\hspace*{-7mm} \widetilde\gamma_E(\alpha,Y) = - \frac{3\alpha_{1}}{5\pi} + \frac{1}{4\pi^2}\, Y_{E}^+ Y_{E} + \frac{1}{2\pi^2}\cdot\frac{117\alpha_{1}^2}{50} + \frac{1}{8\pi^3} Y_{E}^+ Y_{E} \Big(-\frac{3\alpha_{1}}{10} + \frac{3\alpha_{2}}{2} \Big) - \frac{1}{(8\pi^2)^2} \nonumber\\
&&\hspace*{-7mm} \times \bigg[(Y_{E}^+ Y_{E})^2  + Y_{E}^+ Y_{E}\, \mbox{tr} \Big(3 Y_{D}^+ Y_{D} + Y_{E}^+ Y_{E}\Big)\bigg] + O\Big(\alpha^3,\alpha^2 Y^2,\alpha Y^4, Y^6\Big);\\
&&\hspace*{-7mm} \widetilde\gamma_{H_u}(\alpha,Y) = - \frac{3\alpha_{1}}{20\pi} - \frac{3\alpha_{2}}{4\pi} + \frac{3}{8\pi^2}\, \mbox{tr}\Big(Y_{U}^+ Y_{U}\Big) + \frac{1}{2\pi^2}\Big(\frac{207\alpha_{1}^2}{400} + \frac{15\alpha_{2}^2}{16} + \frac{9\alpha_{1}\alpha_{2}}{40}\Big)  \nonumber\\
&&\hspace*{-7mm} +\frac{1}{8\pi^3} \mbox{tr}\Big(Y_{U} Y_{U}^+\Big) \Big(\frac{\alpha_{1}}{5} + 4\alpha_{3} \Big)
-\frac{1}{(8\pi^2)^2}\bigg[\frac{3}{2}\,\mbox{tr}\Big(Y_{D} Y_{D}^+ Y_{U} Y_{U}^+\Big) + \frac{9}{2}\,\mbox{tr}\Big((Y_{U}Y_{U}^+)^2\Big)\bigg]\nonumber\\
&&\hspace*{-7mm} + O\Big(\alpha^3,\alpha^2 Y^2,\alpha Y^4, Y^6\Big);\\
\label{MSSM_Anomalous_Dimentions_Hd_DR}
&&\hspace*{-7mm} \widetilde\gamma_{H_d}(\alpha,Y) = - \frac{3\alpha_{1}}{20\pi} - \frac{3\alpha_{2}}{4\pi} + \frac{1}{8\pi^2}\, \mbox{tr}\Big(3\, Y_{D}^+ Y_{D} + Y_{E}^+ Y_{E}\Big)  + \frac{1}{2\pi^2}\Big(\frac{207\alpha_{1}^2}{400} + \frac{15\alpha_{2}^2}{16} + \frac{9\alpha_{1}\alpha_{2}}{40}\Big) \nonumber\\
&&\hspace*{-7mm}  + \frac{1}{8\pi^3}\mbox{tr}\Big(Y_{E} Y_{E}^+\Big) \cdot \frac{3\alpha_{1}}{10} + \frac{1}{8\pi^3} \mbox{tr}\Big(Y_{D} Y_{D}^+\Big)\Big(-\frac{\alpha_{1}}{10} + 4\alpha_{3}\Big) - \frac{1}{(8\pi^2)^2} \bigg[\frac{3}{2}\mbox{tr}\Big((Y_{E} Y_{E}^+)^2\Big)\nonumber\\
&&\hspace*{-7mm}  + \frac{3}{2}\mbox{tr}\Big(Y_{D} Y_{D}^+ Y_{U} Y_{U}^+\Big) + \frac{9}{2} \mbox{tr}\Big((Y_{D} Y_{D}^+)^2\Big)\bigg]
+ O\Big(\alpha^3,\alpha^2 Y^2,\alpha Y^4, Y^6\Big).
\end{eqnarray}

\noindent
The $\overline{\mbox{DR}}$ expressions for the three-loop MSSM $\beta$-functions (first obtained in \cite{Jack:2004ch}) in this notation take the form

\begin{eqnarray}\label{MSSM_Beta3_DR}
&&\hspace*{-5mm} \frac{\widetilde\beta_3(\alpha,Y)}{\alpha_{3}^2} = - \frac{1}{2\pi} \bigg\{3 -\frac{11\alpha_{1}}{20\pi} -\frac{9\alpha_{2}}{4\pi} -\frac{7\alpha_{3}}{2\pi}
+ \frac{1}{8\pi^2}\, \mbox{tr}\Big(2\, Y_{U}^+ Y_{U} + 2\, Y_{D}^+ Y_{D}\Big) + \frac{1}{2\pi^2}\Big( \frac{851\alpha_{1}^2}{300}\nonumber\\
&&\hspace*{-5mm}  + \frac{27\alpha_{2}^2}{8} - \frac{347\alpha_{3}^2}{24} + \frac{3\alpha_{1}\alpha_{2}}{40} - \frac{11\alpha_{1}\alpha_{3}}{60}
-\frac{3\alpha_{2}\alpha_{3}}{4} \Big) + \frac{1}{8\pi^3}\mbox{tr}\Big(Y_{U} Y_{U}^+\Big)\Big(\frac{11\alpha_{1}}{30} + \frac{3\alpha_{2}}{2} + \frac{13\alpha_{3}}{3} \Big)\nonumber\\
&&\hspace*{-5mm} + \frac{1}{8\pi^3}\mbox{tr}\Big(Y_{D} Y_{D}^+\Big)\Big( \frac{4\alpha_{1}}{15} + \frac{3\alpha_{2}}{2}
+ \frac{13\alpha_{3}}{3}\Big) - \frac{1}{(8\pi^2)^2}\bigg[\, 3\, \mbox{tr}\Big((Y_{U} Y_{U}^+)^2\Big) + 3\, \mbox{tr}\Big((Y_{D} Y_{D}^+)^2\Big)\nonumber\\
&&\hspace*{-5mm} + \frac{9}{2} \Big(\mbox{tr}(Y_{U} Y_{U}^+)\Big)^2
+ \frac{9}{2}\Big(\mbox{tr}(Y_{D} Y_{D}^+)\Big)^2 + \frac{3}{2}\,\mbox{tr}\Big(Y_{E} Y_{E}^+\Big)\,\mbox{tr}\Big(Y_{D} Y_{D}^+\Big) + 2\,\mbox{tr}\Big(Y_{D} Y_{D}^+ Y_{U} Y_{U}^+\Big) \bigg]\bigg\}
\nonumber\\
&&\hspace*{-5mm} + O\Big(\alpha^3,\alpha^2 Y^2, \alpha Y^4, Y^6\Big);\vphantom{\frac{1}{2}}\qquad\\
&&\vphantom{1}\nonumber\\
\label{MSSM_Beta2_DR}
&&\hspace*{-5mm} \frac{\widetilde\beta_2(\alpha,Y)}{\alpha_{2}^2} = - \frac{1}{2\pi} \bigg\{-1  -\frac{9\alpha_{1}}{20\pi}- \frac{25\alpha_{2}}{4\pi} -\frac{6\alpha_{3}}{\pi} + \frac{1}{8\pi^2}\,
\mbox{tr}\Big(3\, Y_{U}^+ Y_{U} + 3\, Y_{D}^+ Y_{D} + Y_{E}^+ Y_{E} \Big) + \frac{1}{2\pi^2}\nonumber\\
&&\hspace*{-5mm} \times\Big( \frac{457\alpha_{1}^2}{200} - \frac{35\alpha_{2}^2}{8} - \frac{11\alpha_{3}^2}{2} - \frac{9\alpha_{1}\alpha_{2}}{40} + \frac{\alpha_{1}\alpha_{3}}{5} - 3\alpha_{2}\alpha_{3}\Big)
+ \frac{1}{8\pi^3}\mbox{tr}\Big(Y_{U} Y_{U}^+\Big)\Big( \frac{29\alpha_{1}}{40} + \frac{33\alpha_{2}}{8}  \nonumber\\
&&\hspace*{-5mm}
+ 4\alpha_{3}\Big) + \frac{1}{8\pi^3}\mbox{tr}\Big(Y_{D} Y_{D}^+\Big)\Big( \frac{11\alpha_{1}}{40}+\frac{33\alpha_{2}}{8} +4\alpha_{3}\Big)
+ \frac{1}{8\pi^3}\mbox{tr}\Big(Y_{E} Y_{E}^+\Big)\Big(\frac{21\alpha_{1}}{40} + \frac{11\alpha_{2}}{8}\Big)
- \frac{1}{(8\pi^2)^2}\nonumber\\
&&\hspace*{-5mm} \times \bigg[\, 6\,\mbox{tr}\Big((Y_{U} Y_{U}^+)^2\Big) + 6\,\mbox{tr}\Big((Y_{D} Y_{D}^+)^2\Big)
+ 2\,\mbox{tr}\Big((Y_{E} Y_{E}^+)^2\Big) + \frac{9}{2}\Big(\mbox{tr}(Y_{U} Y_{U}^+)\Big)^2 + \frac{9}{2}\Big(\mbox{tr}(Y_{D} Y_{D}^+)\Big)^2
\nonumber\\
&&\hspace*{-5mm}
+ \frac{1}{2}\Big(\mbox{tr}(Y_{E} Y_{E}^+)\Big)^2  + 3\, \mbox{tr}\Big(Y_{E} Y_{E}^+\Big)\, \mbox{tr}\Big(Y_{D} Y_{D}^+\Big)
+ 3\,\mbox{tr}\Big(Y_{D} Y_{D}^+ Y_{U} Y_{U}^+\Big) \bigg\} + O\Big(\alpha^3,\alpha^2 Y^2, \alpha Y^4, Y^6\Big);\nonumber\\
&&\vphantom{1}\\
\label{MSSM_Beta1_DR}
&&\hspace*{-5mm} \frac{\widetilde\beta_1(\alpha,Y)}{\alpha_{1}^2} = - \frac{1}{2\pi}\cdot \frac{3}{5} \bigg\{ -11 -\frac{199\alpha_{1}}{60\pi} -\frac{9\alpha_{2}}{4\pi} -\frac{22\alpha_{3}}{3\pi}
+ \frac{1}{8\pi^2} \mbox{tr}\Big(\frac{13}{3} Y_{U}^+ Y_{U} + \frac{7}{3} Y_{D}^+ Y_{D} + 3 Y_{E}^+ Y_{E} \Big) \nonumber\\
&&\hspace*{-5mm}
+ \frac{1}{2\pi^2}\Big( \frac{32117\alpha_{1}^2}{1800} + \frac{27\alpha_{2}^2}{8} - \frac{121\alpha_{3}^2}{18} + \frac{23\alpha_{1}\alpha_{2}}{40} + \frac{137\alpha_{1}\alpha_{3}}{45} + \alpha_{2}\alpha_{3}\Big)
+ \frac{1}{8\pi^3}\mbox{tr}\Big(Y_{U} Y_{U}^+\Big)\Big(\frac{169\alpha_1}{360}\nonumber\\
&&\hspace*{-5mm} + \frac{29\alpha_{2}}{8} + \frac{44\alpha_{3}}{9}\Big)
+ \frac{1}{8\pi^3}\mbox{tr}\Big(Y_{D} Y_{D}^+\Big)\Big( \frac{49\alpha_1}{360} + \frac{11\alpha_{2}}{8} + \frac{32\alpha_{3}}{9}\Big)
+ \frac{1}{8\pi^3}\mbox{tr}\Big(Y_{E} Y_{E}^+\Big)\Big(\frac{27\alpha_1}{40} + \frac{21\alpha_{2}}{8} \Big)
\nonumber\\
&&\hspace*{-5mm}
- \frac{1}{(8\pi^2)^2}\bigg[\,7\,\mbox{tr}\Big((Y_{U} Y_{U}^+)^2\Big) + \frac{9}{2} \mbox{tr}\Big((Y_{D} Y_{D}^+)^2\Big) + \frac{9}{2} \mbox{tr}\Big((Y_{E} Y_{E}^+)^2\Big) + \frac{29}{6}\mbox{tr}\Big(Y_{D} Y_{D}^+ Y_{U} Y_{U}^+\Big)   \nonumber\\
&&\hspace*{-5mm} + \frac{15}{2} \Big(\mbox{tr}(Y_{U} Y_{U}^+)\Big)^2 + 3 \Big(\mbox{tr}(Y_{D} Y_{D}^+)\Big)^2 + 2 \Big(\mbox{tr}(Y_{E} Y_{E}^+)\Big)^2
+ 7\, \mbox{tr}\Big(Y_{E} Y_{E}^+\Big)\,\mbox{tr}\Big(Y_{D} Y_{D}^+\Big) \bigg]\bigg\}\nonumber\\
&&\hspace*{-5mm} + O(\alpha^3,\alpha^2 Y^2, \alpha Y^4, Y^6).\vphantom{\frac{1}{2}}
\end{eqnarray}

\end{document}